\documentclass[10pt,final, twocolumn]{IEEEtran}
\newif\ifFullVersion

\usepackage{cite}
\usepackage{amsmath,amssymb,amsfonts}
\usepackage{graphicx}
\usepackage{textcomp}
\usepackage{xcolor}
\usepackage{tikz}
\usepackage{xcolor}
\usetikzlibrary{positioning, arrows.meta, shapes.geometric, fit, backgrounds, shadows}
\usepackage{multirow}
\usepackage[table,xcdraw]{xcolor}
\usepackage[ruled,linesnumbered,vlined]{algorithm2e}

\usepackage{multirow} 
\usepackage{makecell} 

\definecolor{lightblue}{RGB}{173, 216, 230}
\definecolor{lightorange}{RGB}{255, 218, 185}
\definecolor{lightgreen}{RGB}{144, 238, 144}
\definecolor{lightyellow}{RGB}{255, 255, 224}
\definecolor{lightpink}{RGB}{255, 182, 193}
\definecolor{lightpurple}{RGB}{221, 160, 221}
\definecolor{mediumblue}{RGB}{135, 206, 235}
\definecolor{mediumgreen}{RGB}{152, 251, 152}
\definecolor{darkblue}{RGB}{70, 130, 180}
\definecolor{darkpurple}{RGB}{138, 43, 226}

\tikzset{
    block/.style={
        rectangle, 
        draw, 
        thick, 
        rounded corners=4pt,
        minimum height=2.5cm, 
        minimum width=1.2cm,
        text width=1cm, 
        align=center, 
        font=\tiny\bfseries,
        inner sep=0.5pt,
        text depth=0pt,
        text height=0.2cm
    },
    wideblock/.style={
        rectangle, 
        draw, 
        thick, 
        rounded corners=4pt,
        minimum height=2.5cm, 
        text width=1.8cm,
        align=center, 
        font=\tiny\bfseries,
        inner sep=0.2pt
    },
    smallblock/.style={
        rectangle, 
        draw, 
        thick, 
        rounded corners=3pt,
        minimum height=0.8cm, 
        text width=1.3cm, 
        align=center, 
        font=\scriptsize,
        inner sep=1pt
    },
    circleblock/.style={
        circle,
        draw,
        thick,
        minimum size=0.6cm,
        align=center,
        font=\scriptsize,
        inner sep=1pt
    },
    arrow/.style={-{Stealth[length=2mm, width=1.5mm]}, thick},
    dashedarrow/.style={-{Stealth[length=2mm, width=1.5mm]}, thick, dashed, red},
    estimatorbg/.style={
        fill=lightblue!20, 
        draw=darkblue, 
        rounded corners=6pt,
        inner sep=8pt
    },
    monitorbg/.style={
        fill=lightpurple!15, 
        draw=darkpurple, 
        rounded corners=6pt,
        inner sep=8pt
    }
}

\def\BibTeX{{\rm B\kern-.05em{\sc i\kern-.025em b}\kern-.08em
    T\kern-.1667em\lower.7ex\hbox{E}\kern-.125emX}}

\usepackage{acronym}
\usepackage[belowskip=-10pt,aboveskip=1.8pt]{caption}
\usepackage{standalone}
\usepackage{subcaption}
\usepackage{tikz}
\usepackage{url,enumitem, cite}
\usepackage{verbatim}
\usepackage[bookmarks,colorlinks]{hyperref}
\usepackage{soul}
\usepackage{mathrsfs}
\usepackage[ruled,linesnumbered,vlined]{algorithm2e}
\usepackage{dsfont}
\SetKwInput{KwData}{\textbf{Init}}

\usepackage[all=normal,paragraphs=tight,floats=normal,mathspacing=normal,wordspacing=tight,charwidths=tight,mathdisplays=normal,leading=normal]{savetrees}
\usepackage{pgfplots}
\usepackage{pgfplotstable}
\usetikzlibrary{spy}
\pgfplotsset{compat=newest}
\usepackage{booktabs}

\usepackage{amsmath,amssymb,mathtools}
\usepackage{empheq} 
\newcommand{\myVec}[1]{{\boldsymbol{#1}}}

\newcommand{\mySet}[1]{\mathcal{#1}}

\let\oldnl\nl
\newcommand{\nonl}{\renewcommand{\nl}{\let\nl\oldnl}}

\newcommand{\freqAmbi}[1]{\myVec{a}_{#1}(k)}
\newcommand{\hatfreqAmbi}[1]{\hat{\myVec{a}}_{#1}(k)}

\newcommand{\compressSTFT}[1]{\mathcal{H}(#1)}
\newcommand{\uncompressSTFT}[1]{\mathcal{H}^{-1}(#1)}

\acrodef{edm}[EDM]{elucidating diffusion model}
\acrodef{ai}[AI]{artificial intelligence}
\acrodef{adaln}[AdaLN]{adaptive linear norm}
\acrodef{dl}[DL]{deep learning}
\acrodef{bs}[BS]{base station}
\acrodef{nn}[NN]{neural network}
\acrodef{cnn}[CNN]{convolutional neural network}
\acrodef{mlp}[MLP]{multi-layer perceptron}
\acrodef{snr}[SNR]{signal-to-noise ratio}
\acrodef{awgn}[AWGN]{additive white Gaussian noise} 
\acrodef{ml}[ML]{machine learning} 
\acrodef{sgd}[SGD]{stochastic gradient descent} 
\acrodef{gd}[GD]{gradient descent} 
\acrodef{mse}[MSE]{mean-squared error}
\acrodef{msie}[MSIE]{mean-squared innovation error}
\acrodef{rmse}[RMSE]{root mean squared error}
\acrodef{mspe}[MSPE]{mean-squared periodic error}
\acrodef{rmspe}[RMSPE]{root \ac{mspe}}
\acrodef{mle}[MLE]{maximum likelihood estimation}
\acrodef{snr}[SNR]{signal-to-noise ratio}
\acrodef{admm}[ADMM]{alternating direction method of multipliers}
\acrodef{aoa}[AoA]{Angle of Arrival}
\acrodefplural{aoa}[AoAs]{Angles of Arrival}
\acrodef{em}[EM]{electromagnetic}
\acrodef{cmos}[CMOS]{complementary metal-oxide semiconductor}
\acrodef{ula}[ULA]{uniform linear array}
\acrodef{em}[EM]{Electromagnetic}
\acrodef{doa}[DoA]{direction of arrival}
\acrodef{music}[MUSIC]{MUltiple SIgnal Classification}
\acrodef{esprit}[ESPRIT]{Estimation of signal parameters via rotational invariance techniques}
\acrodef{evd}[EVD]{eigenvalues decomposition}
\acrodef{sps}[SPS]{spatial smoothing}
\acrodef{iid} [i.i.d] {Independent and identically distributed}
\acrodef{ls}[LS]{Least Square}
\acrodef{relu}[ReLu]{Rectified Linear Unit}
\acrodef{crb} [CRB] {Cramér–Rao Bound}
\acrodef{ccrb} [CCRB] {Conditional Cramér–Rao Bound} 
\acrodef{vqvae}[VQ-VAE]{vector quantized variational autoencoder}
\acrodef{rm}[Root-MUSIC]{Root-MUSIC}
\acrodef{esprit}[ESPRIT]{Estimation of Signal Parameters via Rotational Invariance Techniques}
\acrodef{drm}[DR-MUSIC]{Deep Root-MUSIC}
\acrodef{ssn}[SubspaceNet]{Subspace Net}
\acrodef{rssn}[Remote SubspaceNet]{Remote Subspace Net}
\acrodef{mbdl}[Model Based Deep Learning]{Model Based Deep Learning}
 
\acrodef{kf}[KF]{Kalman filter} 
\acrodef{doa}[DOA]{direction-of-arrival} 
\acrodef{sh}[SH]{spherical harmonics} 
\acrodef{sgm}[SGM]{score-based generative model} 
\acrodef{sde}[SDE]{stochastic differential equation} 
\acrodef{au}[AU]{Ambisonics upscaling} 
\acrodef{foa}[FOA]{first-order Ambisonics} 
\acrodef{hoa}[HOA]{high-order Ambisonics} 
\acrodef{stft-sdr}[STFT-SDR]{STFT signal-to-distortion ratio} 
\acrodef{stft-SISDR}[STFT-SISDR]{STFT scale invariant signal-to-distortion ratio}

\acrodef{ve}[VE]{Variance Exploding} 
\acrodef{ode}[ODE]{ordinary differential equation}
\acrodef{dps}[DPS]{Diffusion Posterior Sampling} 
\acrodef{atf}[ATF]{array transfer function} 
\acrodef{sht}[SHT]{Spherical Harmonics Transform} 
\acrodef{arir}[ARIR]{Ambisonics room impulse response} 
\acrodef{ctn}[CTN]{ConvTasNet} 
\acrodef{anova}[ANOVA]{analysis of variance}

\acrodef{tf}[TF]{time-frequency}
\acrodef{stft}[STFT]{short-time Fourier transform}
\acrodef{istft}[ISTFT]{inverse STFT}
\acrodef{ncsn}[NSCN++]{noise-conditioned score-matching network}
\acrodef{rd}[RD]{reverse diffusion}
\acrodef{ald}[ALD]{annealed Langevin dynamics}
\acrodef{mushra}[MUSHRA]{MUltiple Stimuli with Hidden Reference and Anchor}
\acrodef{pwd}[PWD]{plane wave decomposition}
\acrodef{cs}[CS]{compressed sensing}
\acrodef{ild}[ILD]{interaural level difference}
\acrodef{ic}[IC]{interaural coherence}

\acrodef{cs}[CS]{compressed sensing}
\acrodef{sisdr}[SI-SDR]{Signal Invariant Signal Distortion Ratio}
\acrodef{erb}[ERB]{equivalent rectangular bandwidth}
\acrodef{film}[FiLM]{Feature-wise Linear Modulation}
\acrodef{adeups}[ADEUPS]{Ambisonics Diffusion Encoding and Upscaling via Posterior Sampling}
\acrodef{adeps}[ADEPS]{Ambisonics Diffusion Encoding via Posterior Sampling}

\setstcolor{blue}

\begin{document}

\title{Array-Agnostic High-Order Ambisonics Encoding
via Diffusion Posterior Sampling
}

\author{
\IEEEauthorblockN{Amit Milstein, Nir Shlezinger, and Boaz Rafaely}
\thanks{
The preliminary findings of this research were submitted to the International Conference on Acoustics, Speech, and Signal Processing (ICASSP 2027) as the paper \cite{milstein2026array}.  
The authors are with the ECE School Ben-Gurion University of the Negev, Be’er-Sheva, Israel (e-mail: amitmils@post.bgu.ac.il; \{nirshl; br\}@bgu.ac.il). }}

\maketitle

\begin{abstract}
Ambisonics provides a device-independent representation of three-dimensional sound fields; however, extracting these coefficients from practical microphone array recordings remains strongly dependent on array geometry, acoustic transfer functions, and spatial sampling resolution. This pipeline is typically divided into two distinct tasks: encoding, which entails reconstructing the physically resolvable spatial components, and upscaling, which requires generating higher-order coefficients. Model-based approaches often fail on irregular arrays due to ill-conditioned transfer matrices, whereas conventional neural encoders remain strictly tied to the specific acquisition setup seen during training. Furthermore, existing upscaling methods usually assume ideal low-order inputs, ignoring device-specific array artifacts. We present \ac{adeups}, a unified framework that couples an array-independent Ambisonic diffusion prior with an explicit physical acquisition model during inference, enabling zero-shot encoding and upscaling across diverse array configurations. We train our prior using an order-masking strategy and perform inference via score blending, which we demonstrate to be crucial for solving the underlying inverse problem. Through extensive numerical evaluations across various array geometries, we show that \ac{adeups} consistently outperforms task- and array-specific baselines. Finally, we conduct subjective listening tests to confirm the perceptual effectiveness of our approach.
\end{abstract}

\acresetall

 
\section{Introduction}
\label{sec:intro}
Spatial audio conveys the directional and reverberant structure of an acoustic scene and is therefore central to immersive communication, media, and extended-reality applications
\cite{hacihabiboglu2017perceptual}.
Among spatial audio representations, Ambisonics is particularly attractive as it represents the sound field through spherical harmonics independently of the device used to capture it~\cite{zotter2019ambisonics}. In practice, however, obtaining a desired Ambisonics representation from microphone recordings is constrained by the physical acquisition system. A microphone array observes the sound field through its geometry and acoustic transfer functions, such that limited spatial sampling, microphone placement, diffraction, and steering-matrix ill-conditioning introduce frequency- and array-dependent distortions~\cite{rafaely2005analysis}. Moreover, the finite spatial resolution of the array generally allows only a subset of the desired spherical-harmonic orders to be reliably extracted. Consequently, practical Ambisonics acquisition entails two coupled reconstruction tasks: $(i)$ {\em encoding}, which recovers the spatial components observed by the microphone measurements while mitigating acquisition-induced distortions; and $(ii)$ {\em upscaling}, which infers higher-order components that may not be accurately observable by the array. Achieving high-fidelity Ambisonics from practical microphone arrays thus requires addressing both tasks jointly.

Existing approaches have largely addressed these two components of Ambisonics acquisition separately. For {\em encoding}, classical signal-independent methods recover the Ambisonics coefficients by inverting a theoretical or measured array steering model, typically regularized to maintain solution stability~\cite{politis2017comparing,moreau20063d,poletti2005three,bastine2022ambisonics,gayer2025ambisonics}. While applicable to different array configurations given their physical models, their accuracy deteriorates when the corresponding inverse problem is ill-conditioned or spatially underdetermined. Signal-dependent parametric approaches instead estimate a compact spatial description of the acoustic scene and use it to resynthesize the desired Ambisonics representation~\cite{mccormack2022parametric,politis2018compass}. Such methods provide strong physical interpretability, but rely on accurate estimation of spatial parameters and can degrade in reverberant and multi-source environments. Data-driven encoders seek to overcome these limitations by learning the encoding mapping from examples, either by estimating an array-specific linear transform~\cite{heikkinen2024neural,deppisch2026residual} or by directly mapping microphone recordings to Ambisonics coefficients~\cite{qiao2025neural}. Recent approaches improve generalization by conditioning the learned mapping on microphone locations~\cite{heikkinen2025gen} or on measured \acp{atf}~\cite{heikkinen2026beyond}. Nevertheless, the learned mapping remains coupled to the acquisition interface, e.g., through the assumed microphone count or array-dependent training data, limiting its applicability across arbitrary recording configurations.

A complementary line of work addresses {\em Ambisonics upscaling}, where higher-order spatial components are inferred from an available lower-order representation. Classical approaches exploit structural models of the sound field, most commonly through sparse plane-wave decomposition and compressed sensing~\cite{wabnitz2011upscaling,wabnitz2012frequency,epain2013super,routray2020sparse,kentgens2021upscaling}, or through parametric spatial analysis and resynthesis~\cite{politis2015sector,politis2018compass}. These techniques can effectively recover higher-order information when the assumed spatial structure holds, but become less reliable in complex reverberant scenes where the sound field is not well represented by a small number of directional components. More recently, data-driven approaches have learned the low-to-high-order mapping using multi-stage and hierarchical neural architectures~\cite{routray2019deep,gao2022sparse,chatzimoustafa2025higher}, waveform-domain networks~\cite{luo2019conv,nawfalambisonics}, and physics-informed learning~\cite{miotello2024physics}. Generative models have also recently been explored for recovering the inherently underdetermined higher-order components~\cite{zhang2021optimization,milstein2025diffau,picard2026sirup,milstein2026generativelearningambisonicupscaling}. However, these approaches generally operate on already-encoded Ambisonics signals and assume that their observed low-order components provide a reliable representation of the underlying sound field. Consequently, they do not account for the array-dependent distortions, spatial aliasing, and estimation errors introduced during the preceding encoding stage.


While encoding and upscaling can be applied separately, array-dependent distortions and reconstruction errors introduced during encoding can be carried into the upscaling stage, whose input may differ from the ideal Ambisonics signals used for its design or training. 
Rather than learning separate mappings for encoding and upscaling, one can formulate the recovery of an ideal \ac{hoa} representation as a single inverse problem directly from microphone measurements. This motivates a joint approach to encoding and upscaling, recovering the sound field despite partial and distorted observations. Our prior work, \acs{adeps}~\cite{milstein2026array}, adopted this approach for Ambisonic signals via \ac{dps}~\cite{chung2022diffusion}, focusing primarily on refining spatially observable orders, i.e. encoding. Concurrently, Moliner et~al. \cite{moliner2026ambisonics} applied a similar formulation to \acp{arir}.
%
%
However, despite the competitive encoding performance of \acs{adeps}, the recovery of higher-order components degrades drastically as the order increases---both objectively and subjectively. Motivated by this limitation, we propose \emph{\ac{adeups}}, a unified, generative, array-agnostic framework for reconstructing \ac{hoa} representations directly from microphone-array measurements. To this end, we train a diffusion model on ideal \ac{hoa} using a novel \textit{order-masking strategy}, which enforces interchannel dependencies and learns a robust sound-field prior independent of the acquisition hardware. At inference, the microphone configuration is captured by an acoustic forward model that samples the target sound field at the physical  locations and re-encodes the resulting signals into Ambisonics. Embedded within \ac{dps}, this operator, combined with \textit{score blending}, steers the diffusion reverse process to yield the reconstructed \ac{hoa} representation. 
Together, these two contributions allow the same learned prior to compensate for acquisition-induced distortions while inferring spatial components beyond the array's resolvable order. This separation between the learned prior and the physical observation model enables zero-shot operation across microphone geometries and channel counts without retraining. 
Extensive numerical evaluations and subjective listening tests demonstrate its effectiveness across diverse array configurations and reconstruction settings, where it achieves state-of-the-art performance--- surpassing array-specific and semi-agnostic baselines.

The main contributions are as follows:
\begin{itemize}
    \item \textbf{Array-Agnostic Framework:} 
    By utilizing linearly encoded Ambisonics derived from the captured microphone signals, and performing refinement and upscaling entirely within the Ambisonic domain, our generative model decouples the sound-field prior from array geometry, generalizing zero-shot to arbitrary microphone counts and configurations without retraining.
    \item \textbf{Unified Encoding and Upscaling:} We solve Ambisonics encoding and order upscaling within a single inverse problem, eliminating error accumulation and performance bottlenecks inherent to cascaded architectures.
    \item \textbf{Physical Guided Posterior Inference:} By leveraging the physical measurement operator within test-time posterior sampling, the ideal generative prior offers inductive bias against measurement noise, spatial aliasing, and physical array imperfections by constraining solutions to the manifold of ideal \ac{hoa} signals.
    \item \textbf{Extensive Experimental Validation:}
We comprehensively evaluate \ac{adeups} across diverse across diverse microphone geometries and counts, reconstruction orders, considering both encoding and upscaling settings and comparing against dedicated task- and array-specific baselines. In addition to objective spatial and signal-fidelity measures, we conduct subjective listening tests to assess the perceptual quality of the reconstructed Ambisonics signals.
\end{itemize}

The rest of this paper is organized as follows. Section~\ref{sec:System Model and Problem Formulation} introduces the Ambisonic signal model, establishes the general problem formulation, and provides the mathematical foundation for \ac{dps}. Section~\ref{sec:method} details our proposed framework, where we cast the Ambisonic inverse problem explicitly into the \ac{dps} sampling process, describe our order-based masking strategy for intra-order learning, and present our blended score approach alongside a tailored noise scheduler. Section~\ref{sec:numerical_study} presents a comprehensive numerical study evaluating our approach across various array configurations for both encoding and upscaling tasks against task-specific baselines. Section~\ref{sec:listening_exp} reports the results of our subjective listening tests. Finally, Section~\ref{sec: Conclusions} concludes the paper.


Throughout this paper, we denote multi-dimensional quantities using boldface notation, with their dimensions specified upon introduction. 
Calligraphic letters denote sets, e.g., $\mySet{X}$,  with $\mathbb{R}$ and $\mathbb{C}$ being the sets of real and complex numbers,  while  $\mathbb{I}\{\cdot\}$, $\|\cdot\|$, $\|\cdot\|_F$,  $(\cdot)^T$, $(\cdot)^H$, and $(\cdot)^{\dagger}$   are the indicator function, $\ell_2$ norm, Frobenius norm, transpose,  conjugate transpose, and Moore-Penrose pseudo-inverse, respectively.   

\section{Signal Model and Preliminaries}
\label{sec:System Model and Problem Formulation}

\subsection{Signal Model}
\label{subsec:Signal Model}
Ambisonics is a widely used spatial audio representation based on the spherical Fourier transform~\cite{zotter2019ambisonics}. In this framework, a continuous sound pressure is decomposed into a series of \ac{sh} functions of order $n$ and degree $m$, weighted by frequency-dependent Ambisonics coefficients $a_{nm}(k)$, where $k$ denotes the wavenumber. When sampling this sound field using an arbitrary array of $Q$ omnidirectional microphones, where $\myVec{r}_q = (r_q, \theta_q, \phi_q)$ specifies the position of the $q$-th microphone in spherical coordinates \cite{rafaely2015fundamentals}, the acoustic pressure captured by the $q$-th microphone can be modeled as a linear combination of these coefficients~\cite{rafaely2015fundamentals}.
\begin{equation}\label{eq:microphone_acquisition_model}
    \mathrm{p}_q(k) \!=\! \sum_{n=0}^{\infty} \sum_{m=-n}^{n} \!b_n(k, r_q) Y_n^m(\theta_q, \phi_q) a_{nm}(k) \!+\! \nu_q(k).
\end{equation}
In~\eqref{eq:microphone_acquisition_model}, $b_n(k, r_q)$ denotes the order-dependent radial function, $Y_n^m(\theta_q, \phi_q)$ represents the real-valued \ac{sh} function, and $\nu_q(k)$ is zero-mean additive measurement noise, assumed to be uncorrelated with the Ambisonics coefficients and i.i.d.\ across all microphones.

Because the radial functions $b_n(k, r_q)$ decay rapidly beyond the $n > 
\lceil k r_q \rceil$, the infinite series in~\eqref{eq:microphone_acquisition_model} is practically truncated at an effective maximum order $N_{\mathrm{eff}} \ge \left\lceil k \cdot \max_q r_q \right\rceil$~\cite{rafaely2015fundamentals}
. This truncation yields a finite set of $(N_{\mathrm{eff}}+1)^2$ coefficients, enabling a compact matrix representation of the array acquisition process:
\begin{equation}\label{eq:matrix_signal_acquisition}
    \myVec{\mathrm{p}}(k) = \myVec{V}(k) \freqAmbi{N_{\mathrm{eff}}} + \myVec{\nu}(k) \in \mathbb{C}^Q,
\end{equation}
where $\freqAmbi{N_{\mathrm{eff}}} = [a_{00}(k), \ldots, a_{N_{\mathrm{eff}}N_{\mathrm{eff}}}(k)]^T$ stacks the truncated Ambisonics coefficients, $\myVec{\nu}(k) \in \mathbb{C}^Q$ is the array noise vector, and $\myVec{V}(k) \in \mathbb{C}^{Q \times (N_{\mathrm{eff}}+1)^2}$ is the modal steering matrix. This steering matrix factors into
\begin{equation}\label{eq:decomposed_steering_matrix}
    \myVec{V}(k) = \myVec{Y}_{N_{\mathrm{eff}}}\myVec{B}(k),
\end{equation}
where $\myVec{B}(k) \in \mathbb{C}^{(N_{\mathrm{eff}}+1)^2 \times (N_{\mathrm{eff}}+1)^2}$ is a diagonal matrix containing the radial terms and $\myVec{Y}_{N_{\mathrm{eff}}} \in \mathbb{R}^{Q \times (N_{\mathrm{eff}}+1)^2}$ evaluates the spherical harmonics at the microphone coordinates.

\subsection{Problem Formulation}
\label{sec:ProblemFormulation}
 Ambisonics reconstruction~\cite{poletti2005three} aims to estimate the \ac{sh} coefficients from discrete microphone array measurements~\eqref{eq:matrix_signal_acquisition}. Accordingly, our objective is to reconstruct a desired order-$N_p \le N_{\text{eff}}$ Ambisonics representation directly from the microphone-array measurements in~\eqref{eq:matrix_signal_acquisition}. We represent such a reconstruction by an operator $\mathrm{R}$ satisfying
\begin{equation}
    \hatfreqAmbi{N_p} = \mathrm{R}\left( \myVec{\mathrm{p}}(k), \myVec{V}(k) \right),
    \label{eq:encoding}
\end{equation}
where $\hatfreqAmbi{N_p}\in\mathbb{C}^{(N_p+1)^2}$ denotes the reconstructed Ambisonics coefficients. Unlike conventional encoding, the target order $N_p$ need not be restricted to the spatial order that can be reliably recovered through direct inversion of the array model. Consequently,~\eqref{eq:encoding} encompasses both {\em encoding}, namely the recovery of spatial components supported by the microphone observations, and {\em upscaling}, namely the inference of higher-order components that are insufficiently observed by the array. 

 By~\eqref{eq:decomposed_steering_matrix}, the mapping \eqref{eq:encoding} can be formulated  as an inverse problem~\cite{politis2017comparing, heikkinen2024neural}. However, recovering $\freqAmbi{N_p}$ from~\eqref{eq:encoding} gives rise to several challenges:
\begin{enumerate}[label=$C_\arabic*$, leftmargin=*,series=challenges]
   \item \label{item:c1} \textbf{Noise Amplification:} At low frequencies ($k \to 0$), the decay in $\myVec{B}(k)$ at higher orders $n$ renders direct matrix inversion severely ill-posed, leading to extreme microphone noise amplification.
\item \label{item:c2} \textbf{Non-Uniform Sensitivity:} Irregular or non-ideal spatial arrangements of microphones degrade the condition number of $\myVec{Y}_{N_{\mathrm{eff}}}$, causing direction- and channel-dependent reconstruction errors.
\item \label{item:c3} \textbf{High-Frequency Spatial Aliasing:}  At frequencies for which the microphone array does not sufficiently sample the sound field, spatial aliasing introduces frequency- and geometry-dependent distortions into the recoverable Ambisonics coefficients.
\item \label{item:c4} \textbf{Underdetermined High-Order Recovery:} When the desired Ambisonics order exceeds the spatial dimensionality supported by the microphone observations, e.g., $Q<(N_p+1)^2$, direct reconstruction becomes underdetermined. Conventional encoding must therefore restrict the recovered order, leaving higher-order spatial information unresolved.
    \item \label{item:c5} \textbf{Array-Invariance:} 
    The encoder should be applicable across array   geometries and microphone counts $Q$.
\end{enumerate} 

Although \ref{item:c1}--\ref{item:c3} represent inherent physical degradations in Ambisonic encoding while \ref{item:c4} pertains to Ambisonics upscaling, both stem from the same underlying inverse problem. Our objective is thus to establish a unified spatial mapping operator that addresses these challenges across diverse array configurations (\ref{item:c5}). To this end, we assume access during training to ideal, array-agnostic $N_p$-order Ambisonic coefficients, formalized as the dataset
\begin{equation}
    \label{eqn:Dataset}
    \mySet{D} = \big\{\myVec{a}_{N_p}^{(i)} \big\}_{i=1}^{|\mySet{D}|}.
\end{equation}

\subsection{Preliminaries}\label{ssec:Preliminaries}
Here we review some basic preliminaries needed to formulate our proposed \ac{adeups} in the following section. These include conventional linear encoding, as well as some basics in generative diffusion modelling and sampling.  

\subsubsection{Linear Encoder}
Conventional linear encoding tackles the encoding component of~\eqref{eq:encoding} by recovering Ambisonics coefficients directly from the microphone measurements through a frequency-dependent linear operator. For a prescribed order $N\leq N_{\mathrm{eff}}$, it forms 
\begin{equation}
    \hatfreqAmbi{N}
    =\myVec{E}_{N}(k)\myVec{\mathrm{p}}(k).
\end{equation}
Let $\myVec{V}_{N}(k)$ contain the columns of $\myVec{V}(k)$ in \eqref{eq:matrix_signal_acquisition} associated with Ambisonics coefficients up to order $N$. A common encoding rule is its Tikhonov-regularized pseudoinverse~\cite{moreau20063d,poletti2005three}:
\begin{equation}\label{eq:linear_encoder}
\begin{aligned}
\myVec{E}_{N}(k)
={}&\myVec{V}_{N}^{H}(k)
\big[
\myVec{V}_{N}(k)
\myVec{V}_{N}^{H}(k)
+\zeta^2\myVec{I}_Q
\big]^{-1},
\end{aligned}
\end{equation}
where   $\zeta^2$ controls the trade-off between inversion accuracy and noise amplification. While this construction applies to any known steering matrix and thus accommodates arbitrary geometries, its output remains constrained by the rank and conditioning of $\mathbf{V}_N(k)$. Consequently, the  maximum achievable order is bounded by $N_{\text{enc}} = \lfloor\sqrt{Q}-1\rfloor$, which, under \ref{item:c4},   cannot upscale to recover the \ac{hoa} if $N_{\text{enc}} < N_{\text{eff}}$. Increasing $\zeta$ suppresses unstable modes at the expense of model mismatch and spatial bias, while decreasing it amplifies microphone noise. 
Linear encoding consequently remains also sensitive to challenges \ref{item:c1}--\ref{item:c3}.

\subsubsection{Diffusion Models}
Diffusion models map a complex data distribution $p_{\text{data}}$ to a tractable prior distribution (typically Gaussian) along a continuous trajectory. In the forward process, a clean sample $\myVec{x}_0 \sim p_{\text{data}}$ is progressively corrupted across time $t \in [0, T_{\max}]$ according to
\begin{equation}\label{eq:edm_forward}
    \myVec{x}_t = \myVec{x}_0 + \sigma(t)\myVec{\epsilon}, \qquad \myVec{\epsilon} \sim \mathcal{N}(\myVec{0}, \myVec{I}),
\end{equation}
where $\sigma(t)$ defines the noise schedule. To generate samples, this process is reversed by solving the reverse-time \ac{sde}~\cite{song2020score_sde}
\begin{equation}\label{eq:reverse_sde}
    \mathrm{d}\myVec{x}_t = \left[ f(\myVec{x}_t, t) - g(t)^2 \nabla_{\myVec{x}_t} \log p_t(\myVec{x}_t) \right] \mathrm{d}t + g(t) \, \mathrm{d}\myVec{\bar{w}},
\end{equation}
where $f(\myVec{x}_t, t)$ is referred to as the drift term, $g(t)$ is the diffusion coefficient, $\myVec{\bar{w}}$ is a time-reversed Wiener process, and $p_t(\myVec{x}_t)$ represents the marginal density of the corrupted state $\myVec{x}_t$ at time $t$. The gradient $\nabla_{\myVec{x}_t} \log p_t(\myVec{x}_t)$ defines the marginal score function.

In practice, a \ac{nn} $s_{\rm prior}(\myVec{x}_t, \sigma(t);\myVec{\theta})$ 
is often trained to estimate this score function. Under the \ac{ve} formulation of \acp{edm}~\cite{karras2022elucidating}, the drift term is removed ($f = 0$) and the diffusion parameter is set to $g(t) = \sqrt{2t}$, which directly yields $\sigma(t) = t$. \acp{edm} precondition the \ac{nn} as a denoiser $D_\myVec{\theta}(\myVec{x}_\sigma, \sigma)$ according to
\begin{equation}\label{eq:edm_preconditioning} 
D_\myVec{\theta}(\myVec{x}_t, \sigma) \!=\! c_{\mathrm{skip}}(\sigma)\myVec{x}_t \!+\! c_{\mathrm{out}}(\sigma) F_\myVec{\theta}\!\left( c_{\mathrm{in}}(\sigma)\myVec{x}_t, c_{\mathrm{noise}}(\sigma) \right), 
\end{equation}
where $F_\myVec{\theta}$ is the model to be trained, and the coefficients are given by 
\begin{subequations}
    \label{eq:standard_edm_coefficients}
\begin{align}
c_{\mathrm{in}}(\sigma) &= (\sigma^2 + \sigma_{\mathrm{data}}^2)^{-1/2}, \\
c_{\mathrm{skip}}(\sigma) &= \frac{\sigma_{\mathrm{data}}^2}{\sigma^2 + \sigma_{\mathrm{data}}^2}, \\
c_{\mathrm{out}}(\sigma) &= \frac{\sigma \cdot \sigma_{\mathrm{data}}}{\sqrt{\sigma^2 + \sigma_{\mathrm{data}}^2}}, \\
c_{\mathrm{noise}}(\sigma) &= \tfrac{1}{4} \ln(\sigma). 
\end{align}
\end{subequations}
Here, $\sigma_{\mathrm{data}}$ represents the standard deviation of the data distribution, while $c_{\mathrm{noise}}(\sigma)$ provides a log-scaled noise embedding. These coefficients normalize the network inputs and output targets across noise levels while preserving a skip connection to the data manifold when $\sigma \to 0$.

Via Tweedie's formula~\cite{robbins1992empirical}, the marginal score function is related to the ideal denoiser by
\begin{equation}\label{eq:tweedie}
\nabla_{\myVec{x}_t} \log p_t(\myVec{x}_t) \approx s_{\rm prior}(\myVec{x}_t, \sigma(t);\myVec{\theta}) = \frac{D_\myVec{\theta}(\myVec{x}_t, \sigma(t)) - \myVec{x}_t}{\sigma(t)^2}.
\end{equation}
Sample generation can thus be executed  by solving the probability flow \ac{ode}~\cite{song2020score}
\begin{align}
     \mathrm{d}\myVec{x}_t &= -\sigma(t) \nabla_{\myVec{x}_t} \log p_t(\myVec{x}_t)\mathrm{d}\sigma(t) \notag \\&\approx -\sigma(t)s_{\rm prior}(\myVec{x}_t, \sigma(t);\myVec{\theta})\mathrm{d}\sigma(t), 
     \label{eq:ODE}
\end{align}
or stochastically by integrating an equivalent \ac{sde} solver.

\subsubsection{Diffusion Posterior Sampling}
\label{ssec:dps_prelim}
Rather than generating samples, this approach can be used to solve inverse problems. Consider an unknown clean sample $\myVec{x}_0$ is observed via
\begin{equation}\label{eq:generic_inverse_problem}
    \myVec{y} = \mathcal{A}(\myVec{x}_0) + \myVec{\epsilon}_y, \qquad \myVec{\epsilon}_y \sim \mathcal{N}(\myVec{0}, \sigma_y^2 \myVec{I}),
\end{equation}
where $\mathcal{A}$ denotes a known, potentially nonlinear degradation operator. Sampling from the posterior distribution $p(\myVec{x}_0 | \myVec{y})$ can be achieved by substituting the posterior score function into the probability flow ODE in~\eqref{eq:ODE}. By applying Bayes' rule, the posterior score decomposes into
\begin{equation}\label{eq:bayes_decomposition}
    \nabla_{\myVec{x}_t} \log p_t(\myVec{x}_t | \myVec{y}) \!=\! \nabla_{\myVec{x}_t} \log p_t(\myVec{x}_t) \!+\! \nabla_{\myVec{x}_t} \log p_t(\myVec{y} | \myVec{x}_t),
\end{equation}
where the first term is supplied by the unconditional diffusion prior and the second term corresponds to the likelihood score which we denote as $\myVec{s}_{\text{LH}}(\myVec{y} \mid \myVec{x}_t)= \nabla_{\myVec{x}_t} \log p_t(\myVec{y} | \myVec{x}_t)$. 

Because the exact likelihood $p_t(\myVec{y} | \myVec{x}_t)$ is intractable for intermediate states $t > 0$, it is commonly approximated via the estimated target $\hat{\myVec{x}}_0 = D_\myVec{\theta}(\myVec{x}_t, \sigma(t))$~\cite{chung2022diffusion, moliner2023solving}:
\begin{align} 
\nabla_{\myVec{x}_t} \log p_t(\myVec{y} \mid \myVec{x}_t) 
&\approx \nabla_{\myVec{x}_t} \log p(\myVec{y} \mid \hat{\myVec{x}}_0) \notag \\
&\stackrel{(a)}{\propto}  -\eta(t) \nabla_{\myVec{x}_t} \left\| \myVec{y} - \mathcal{A}(\hat{\myVec{x}}_0) \right\|_2^2,
\label{eq:likelihood_guidance}
\end{align}
where $(a)$ follows from the Gaussian observation model in \eqref{eq:generic_inverse_problem} and $\eta(t)$ is a time-dependent weight controlling the guidance strength. Substituting this approximation together with \eqref{eq:tweedie} into the probability flow \ac{ode} yields the guided generative process:
\begin{align}
     \mathrm{d}\myVec{x}_t =& -\sigma(t) \Big[s_{\rm prior}(\myVec{x}_t, \sigma(t);\myVec{\theta})\notag \\
     &\quad-\eta(t) \nabla_{\myVec{x}_t} \left\| \myVec{y} - \mathcal{A}(\hat{\myVec{x}}_0)) \right\|_2^2\Big]\mathrm{d}\sigma(t).
     \label{eq:dps_ode}
\end{align}
This gradient steering pushes the sampling trajectory toward reconstructions that remain both realistic under the prior and faithful to the observation $\myVec{y}$. 

\section{ADEUPS}
\label{sec:method}

We next introduce \ac{adeups}, which formulates the inverse problem in the Ambisonics domain to enable zero-shot encoding and upscaling for arbitrary array configurations using a generative prior. Subsection~\ref{ssec:dps_formulation_AE} frames this inverse problem within the \ac{dps} framework, Subsection~\ref{ssec:training} outlines our training scheme with order masking and conditioning, and Subsection~\ref{ssec:inference} details the inference pipeline with scheduled score blending, followed by a discussion in Subsection~\ref{ssec:discussion}. All signals are defined in the \ac{stft} domain; time-frame and frequency-bin indices are omitted for brevity.

\begin{figure*}
    \centering
    \includegraphics[width=1\linewidth]{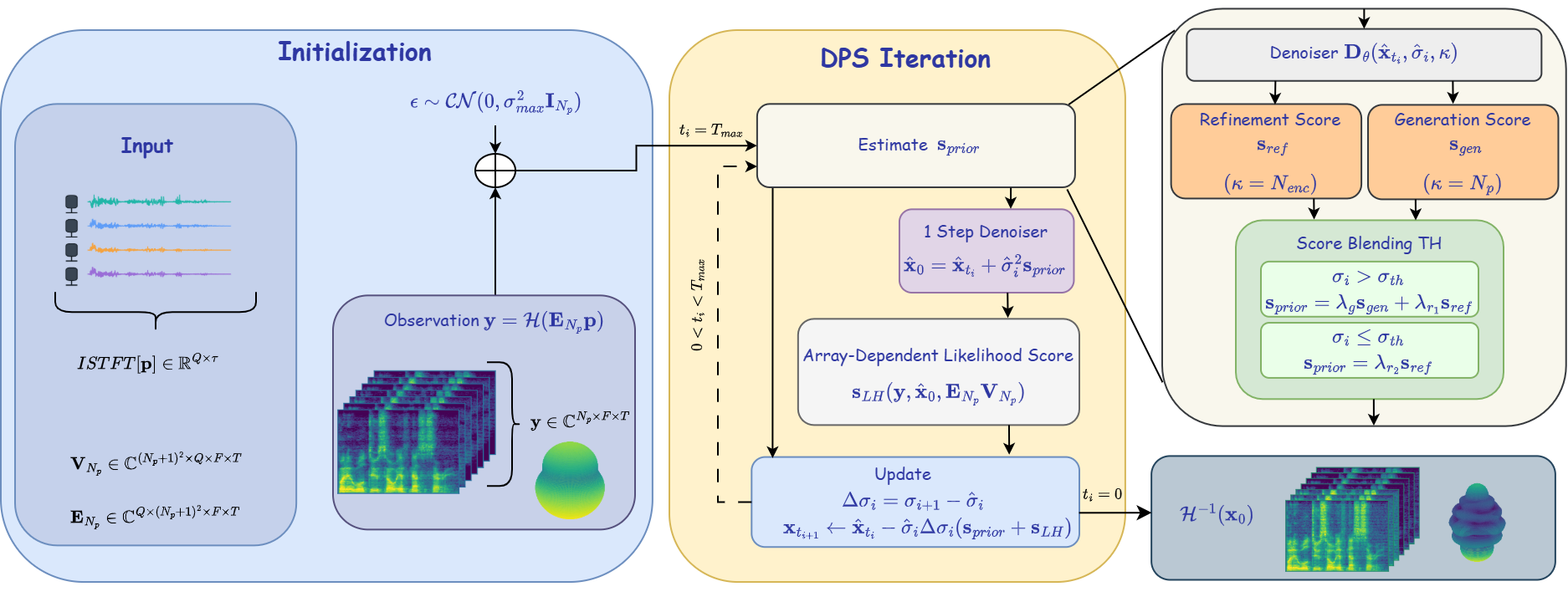}
    \caption{ADEUPS Pipeline}
    \label{fig:adeups_diagram}
\end{figure*}

\subsection{DPS Formulation for Ambisonics Reconstruction}
\label{ssec:dps_formulation_AE}
Recall that due to \ref{item:c4},  the desired Ambisoncs order, which is the order of the Ambisonics prior, holds
 $N_p > N_{\text{enc}}$. Similar to \cite{gerkmann2010empirical,richter2023speech}, we define our prior in the magnitude-compressed \ac{stft} domain. We denote this element-wise compression operator as $\compressSTFT{z} = \beta |z|^{\alpha} e^{i \angle(z)}$, where $\alpha, \beta > 0$ are hyperparameters and $\angle(z)$ denotes the phase angle.

We formulate our inverse problem directly in the Ambisonics domain by defining the observation vector $\myVec{y}$ as the compressed encoded signal of order $N_p$ (matching the prior order). Specifically, writing the observations of \eqref{eq:generic_inverse_problem} as $\myVec{y} = \compressSTFT{\myVec{E}_{N_p}\myVec{\mathrm{p}}} + \myVec{\epsilon}_y$ and substituting the signal acquisition model~\eqref{eq:matrix_signal_acquisition} yields:
\begin{equation}\label{eq:inverse_problem}
    \myVec{y} = \compressSTFT{\myVec{E}_{N_p}\myVec{V}_{N_p}\uncompressSTFT{\myVec{x}_0}} + \myVec{\epsilon}_y.
\end{equation}
Accordingly, we define the forward  operator $\mathcal{A}(\cdot)$ in \eqref{eq:generic_inverse_problem}, which maps the ideal compressed $N_p$ order sample $\myVec{x}_0$ to the corresponding observation domain, as
\begin{equation}\label{eq:degradation_operator}
    \mathcal{A}(\myVec{x}_0) = \compressSTFT{\myVec{E}_{N_p}\myVec{V}_{N_p}\uncompressSTFT{\myVec{x}_0}}.
\end{equation}
This operator explicitly captures the sequence of linear STFT decompression, spatial sampling by microphones, Ambisonics linear encoding, and STFT compression. By modeling these array-specific degradation artifacts~(\ref{item:c1}--\ref{item:c4}) exclusively within the forward pipeline at inference time, the resulting \ac{dps} formulation, once properly trained, suppors zero-shot generalization across arbitrary microphone geometries~(\ref{item:c5}).

\subsection{Inference}
\label{ssec:inference}

We next describe the inference procedure assuming that the parameters $\myVec{\theta}$ of the order-conditioned denoiser $D_\myVec{\theta}(\myVec{x}_t,\sigma(t),\kappa)$ are given; their training is discussed in Subsection~\ref{ssec:training}. The conditioning variable $\kappa$ specifies the highest Ambisonics order provided to the denoiser as trusted context. For an array with $Q$ microphones, we set this order to $N_{\mathrm{enc}}=\lfloor\sqrt{Q}-1\rfloor$.

Recall that \ac{dps} reconstructs the unknown signal by replacing the unconditional score in the reverse diffusion dynamics with the posterior score in~\eqref{eq:bayes_decomposition}. Accordingly, \ac{adeups} inference follows~\eqref{eq:dps_ode} using
\begin{equation}
    \myVec{s}_{\mathrm{post}}     (\myVec{x}_t,\sigma(t),\myVec{y})     = \myVec{s}_{\mathrm{prior}} (\myVec{x}_t,\sigma(t);\myVec{\theta}) + \myVec{s}_{\mathrm{LH}} (\myVec{y}\mid\myVec{x}_t),
    \label{eq:adeups_posterior_score}
\end{equation}
where the prior score promotes consistency with the learned distribution of ideal Ambisonics signals, while the likelihood score enforces consistency with the array observation through the physical forward operator~\eqref{eq:degradation_operator}. We next specify these two terms.

\smallskip
\noindent
{\bf Order-conditioned prior score:}
A single full-order diffusion score does not explicitly distinguish between Ambisonics components that can be supported by the array and higher-order components that must be inferred from their lower-order context. We therefore exploit two complementary operating modes of the same order-conditioned denoiser. Evaluating it with $\kappa=N_{\mathrm{enc}}$ yields the \emph{generation score}
\begin{equation}
    \myVec{s}_{\mathrm{gen}} (\myVec{x}_t,\sigma(t);\myVec{\theta}) =
    \frac{ D_\myVec{\theta} (\myVec{x}_t,\sigma(t),\kappa=N_{\mathrm{enc}}) - \myVec{x}_t} {\sigma(t)^2},
    \label{eq:score_gen}
\end{equation}
which uses the lower-order Ambisonics components as trusted context for generating the remaining \ac{hoa} components. 
Evaluating the denoiser with $\kappa=N_p$ gives the \emph{refinement score}
\begin{equation}
    \myVec{s}_{\mathrm{ref}} (\myVec{x}_t,\sigma(t);\myVec{\theta}) =
    \frac{ D_\myVec{\theta} (\myVec{x}_t,\sigma(t),\kappa=N_p) - \myVec{x}_t} {\sigma(t)^2},
    \label{eq:score_ref}
\end{equation}
which operates on the complete order-$N_p$ representation and promotes full-order denoising and refinement. The mechanism by which these two behaviors are induced during training is detailed in Subsection~\ref{ssec:training}.

The relative importance of generation and refinement varies along the reverse diffusion trajectory. At high noise levels, the higher-order structure has not yet been established, making generation from the trusted lower-order context particularly important. At low noise levels, the reconstructed \ac{hoa} structure is already largely formed, and continued generative guidance can disturb its fine details. We therefore employ score blending~\cite{liu2022compositional} and define
\begin{align}
&\myVec{s}_{\mathrm{prior}} (\myVec{x}_t,\sigma(t);\myVec{\theta}) \notag \\
&\quad= 
\begin{cases}
\!\begin{aligned}
&\lambda_g\,\myVec{s}_{\mathrm{gen}} (\myVec{x}_t,\sigma(t);\myVec{\theta}) \\[-0.5ex]
&\quad+ \lambda_{r_1}\,\myVec{s}_{\mathrm{ref}}  (\myVec{x}_t,\sigma(t);\myVec{\theta}),
\end{aligned}
& \sigma(t)>\sigma_{\mathrm{th}}, \\[2.5ex]
\lambda_{r_2}\,\myVec{s}_{\mathrm{ref}}(\myVec{x}_t,\sigma(t);\myVec{\theta}),
& \sigma(t)\leq\sigma_{\mathrm{th}},
\end{cases}
\label{eq:dual_score_schedule}
\end{align}
where $\lambda_g,\lambda_{r_1},\lambda_{r_2}\geq0$ control the relative score contributions, and $\sigma_{\mathrm{th}}$ is a threshold hyperparameter which is set empirically. Thus, the early stages of sampling jointly establish higher-order structure and refine the current Ambisonics estimate, whereas the final stages (where the noise level is below the threshold) focus on full-order refinement.

\smallskip
\noindent
{\bf Array-dependent likelihood score:}
Having formed the prior score, we instantiate the likelihood term of the generic \ac{dps} construction in~\eqref{eq:likelihood_guidance}. Following Tweedie's relation in~\eqref{eq:tweedie}, the blended score provides the clean-signal proxy
\begin{equation}
    \tilde{\myVec{x}}_0 = \myVec{x}_t + \sigma(t)^2 \myVec{s}_{\mathrm{prior}} (\myVec{x}_t,\sigma(t);\myVec{\theta}).
    \label{eq:adeups_clean_proxy}
\end{equation}
The corresponding likelihood score is
\begin{equation}
    \myVec{s}_{\mathrm{LH}} (\myVec{y}\mid\myVec{x}_t) = -\eta(t) \nabla_{\myVec{x}_t} \left\| \myVec{y} - \mathcal{A}(\tilde{\myVec{x}}_0) \right\|_2^2,
    \label{eq:adeups_likelihood_score}
\end{equation}
where $\mathcal{A}(\cdot)$ is defined in \eqref{eq:degradation_operator}. Following \cite{kim2022guided,moliner2023solving}, we normalize the guidance strength according to
\begin{equation}
    \eta(t) = \frac{\eta'} { \sigma(t) \left\| \nabla_{\myVec{x}_t} \left\| \myVec{y} - \mathcal{A}(\tilde{\myVec{x}}_0) \right\|_2^2 \right\|_2 },
    \label{eq:adeups_guidance_weight}
\end{equation}
where $\eta'>0$ is a constant scaling parameter. Consequently, the specific microphone configuration affects the posterior trajectory only through $\mathcal{A}(\cdot)$ in the likelihood score, while the learned prior score remains independent of the acquisition array.

\smallskip
\noindent
{\bf Posterior sampling:}
Substituting~\eqref{eq:dual_score_schedule} and \eqref{eq:adeups_likelihood_score} into the \ac{dps} dynamics in~\eqref{eq:dps_ode} yields the \ac{adeups} reverse trajectory. We discretize it over $M$ steps using the $\rho$-spaced noise schedule of~\cite{karras2022elucidating},
\begin{equation}
\label{eq:edm_noise_schedule}
    \sigma_i = \left( \sigma_{\max}^{1/\rho} + \frac{i}{M-1} \left( \sigma_{\min}^{1/\rho} - \sigma_{\max}^{1/\rho} \right) \right)^\rho,
\end{equation}
for $i=0,\ldots,M-1$, with $\sigma_M=0$. We initialize the trajectory around the encoded observation as
\begin{equation}
    \myVec{x}_{T_{\max}} \sim \mathcal{CN} \left( \myVec{y}, \sigma_{\max}^2\myVec{I} \right),
\end{equation}
where $\myVec{y}=\compressSTFT{\myVec{E}_{N_p}\myVec{\mathrm{p}}}$. We then use the first-order Euler sampler with stochastic churn~\cite{karras2022elucidating}. Denoting the churn-perturbed state and noise level at iteration $i$ by $\hat{\myVec{x}}_{t_i}$ and $\hat{\sigma}_i$, respectively, the posterior scores are evaluated at $(\hat{\myVec{x}}_{t_i},\hat{\sigma}_i)$ and the state is updated as
\begin{equation}
\begin{aligned}
    \myVec{x}_{t_{i+1}} = \hat{\myVec{x}}_{t_i} - \hat{\sigma}_i (\sigma_{i+1}-\hat{\sigma}_i) \left( \myVec{s}_{\mathrm{prior},i} +
        \myVec{s}_{\mathrm{LH},i} \right).
\end{aligned}
\label{eq:adeups_sampling_update}
\end{equation}
After the final sampling step, applying $\uncompressSTFT{\cdot}$ yields the reconstructed order-$N_p$ Ambisonics coefficients. The complete inference procedure, including the stochastic churn operations, is summarized in Algorithm~\ref{alg:adeups} and illustrated in Figure~\ref{fig:adeups_diagram}.

\begin{algorithm}[t]
\caption{\ac{adeups} Inference}
\label{alg:adeups}
\small
\DontPrintSemicolon
\SetKwInOut{Input}{Input}
\setlength{\abovedisplayskip}{0pt}
\setlength{\belowdisplayskip}{0pt}

\Input{Signals $\myVec{\mathrm{p}} $, Steering matrix $\myVec{V}$, Denoiser $D_\myVec{\theta}(\cdot)$,\\ Hyperparameters $\{\lambda_g, \lambda_{r_1}, \lambda_{r_2}, \eta', \sigma_{\rm th}\}$,\\ Chrun parameters $S_{\rm churn}, S_{\rm noise}, S_{\min}, S_{\max}$}

Set $N_{\mathrm{enc}}=\lfloor\sqrt{Q}-1\rfloor$ and $\myVec{E}_{N_p}$ via~\eqref{eq:linear_encoder} \\
Compute $\myVec{y} = \compressSTFT{\tilde{\myVec{E}}_{N_p}\myVec{\mathrm{p}}}$ \\
Sample $\myVec{x}_{T_{\max}} \sim \mathcal{CN}(\myVec{y}, \sigma_{\max}^2 \myVec{I}_{N_p})$ \\

\For{$i = 0$ \KwTo $M-1$}{
    \tcc{Stochastic Churn sampling}
    $\gamma_i \leftarrow \min \left( \frac{S_{\text{churn}}}{M}, \, \sqrt{2} - 1 \right) \cdot \mathbb{I}\{\sigma_i \in [S_{\text{min}}, S_{\text{max}}]\}$ \\
    $\hat{\sigma}_i \leftarrow (1 + \gamma_i) \, \sigma_i$ \\ Sample $\myVec{\epsilon}_i \sim \mathcal{CN}(\myVec{0}, S_{\text{noise}}^2 \myVec{I})$ \\
    $\hat{\myVec{x}}_{t_i} \leftarrow \myVec{x}_{t_i} + \sqrt{\hat{\sigma}_i^2 - \sigma_i^2} \, \myVec{\epsilon}_i$ \\
    \tcc{Calculate refinement score}
    $\myVec{s}_{\text{ref}} \leftarrow \big( D_\myVec{\theta}(\hat{\myVec{x}}_{t_i}, \hat{\sigma}_i, \kappa=N_p) - \hat{\myVec{x}}_{t_i}\big) / \hat{\sigma}_i^2$ \\
    \tcc{Blend score if over threshold}
    \eIf{$\sigma_i > \sigma_{\text{th}}$}{
        $\myVec{s}_{\text{gen}} \leftarrow \big( D_\myVec{\theta}(\hat{\myVec{x}}_{t_i}, \hat{\sigma}_i, \kappa=N_{\text{enc}}) - \hat{\myVec{x}}_{t_i}\big) / \hat{\sigma}_i^2$ \\
        $\myVec{s}_{\text{prior}} \leftarrow \lambda_g\myVec{s}_{\text{gen}} + \lambda_{r_1}\myVec{s}_{\text{ref}}$
    }{
        $\myVec{s}_{\text{prior}} \leftarrow \lambda_{r_2}\myVec{s}_{\text{ref}}$
    }
    \tcc{Perform DPS step}
    $\tilde{\myVec{x}}_0 \leftarrow \hat{\myVec{x}}_{t_i} + \hat{\sigma}_i^2 \myVec{s}_{\text{prior}}$ \\
    $\myVec{s}_{\text{LH}} \leftarrow -\eta(t_i) \nabla_{\hat{\myVec{x}}_{t_i}} \|\myVec{y} - \mathcal{A}(\tilde{\myVec{x}}_0)\|_2^2$ \\
    Update $\myVec{x}_{t_{i+1}} \leftarrow \hat{\myVec{x}}_{t_i} - \hat{\sigma}_i(\sigma_{i+1} - \hat{\sigma}_i) \left[ \myVec{s}_{\text{prior}} + \myVec{s}_{\text{LH}} \right]$
}
\KwRet{$\uncompressSTFT{\myVec{x}_0}$}
\end{algorithm}

\subsection{Order-Masking Prior Training}
\label{ssec:training}

The inference procedure in Subsection~\ref{ssec:inference} assumes access to an order-conditioned denoiser $D_\myVec{\theta}(\myVec{x}_t,\sigma(t),\kappa)$ that can operate both as a full-order refinement prior and as a conditional generator from a prescribed trusted order $\kappa$. We next describe how these complementary behaviors are induced during training. Training is carried out exclusively using ideal order-$N_p$ Ambisonics signals from the dataset~\eqref{eqn:Dataset}; no microphone signals, steering matrices, or array-dependent encoding operators are used. The resulting learned prior is therefore independent of the acquisition configuration.

To expose the denoiser to different amounts of available spatial information, we randomly mask complete Ambisonics orders during training. Let $c\in\{0,\ldots,(N_p+1)^2-1\}$ denote the Ambisonics channel index and let $\ell(c)=\lfloor\sqrt{c}\rfloor$ denote its corresponding \ac{sh} order. For each training sample, we draw a trusted order $\kappa\in\{1,\ldots,N_p\}$ according to a categorical distribution with probabilities $p_\kappa$, and define
\begin{equation}
    q_\ell  = \mathbb{I}\{\ell\leq\kappa\},
    \qquad
    \ell=0,\ldots,N_p,
    \label{eq:order_trust}
\end{equation}
with $\mathbb{I}$ the indicator function and with the corresponding channel-wise mask $\bar{\myVec{q}}$ whose entries are $[\bar{\myVec{q}}]_c=q_{\ell(c)}$. Thus, all  components up to order $\kappa$ are treated as trusted, while those of orders $\ell>\kappa$ are regarded as unavailable.

Given an ideal sample $\myVec{x}_0$, a noise level $\sigma(t)$, and $\myVec{\epsilon}\sim\mathcal{CN}(\myVec{0},\myVec{I})$, we construct the input to the denoiser channel-wise as
\begin{equation}
    \tilde{\myVec{x}}_{t}  = {\rm diag}\{\bar{\myVec{q}}\} \myVec{x}_{0} + \sigma(t)\myVec{\epsilon}.
    \label{eq:masked_noising}
\end{equation}
Consequently, trusted channels follow the standard diffusion corruption process in~\eqref{eq:edm_forward}, whereas channels above order $\kappa$ contain only noise. The training target, however, always remains the complete clean order-$N_p$ Ambisonics signal $\myVec{x}_0$. Hence, when $\kappa<N_p$, successful denoising requires the network to infer the masked higher-order components from the available lower-order context, thereby learning the inter-order dependencies required by the generation score in~\eqref{eq:score_gen}.
For $\kappa=N_p$, no channels are masked and the model reduces to full-order denoising, as required by the refinement score in~\eqref{eq:score_ref}.

The standard \ac{edm} preconditioning in \eqref{eq:edm_preconditioning} must also be adapted to this masking operation. In particular, directly retaining the skip connection on a masked channel would propagate its pure-noise input to the denoiser output. We therefore define the order-conditioned denoiser as
\begin{align}
D_{\myVec{\theta}} (\tilde{\myVec{x}}_t,\sigma(t),\kappa) = &{\rm diag}\{\myVec{c}_{\mathrm{skip}}(\sigma(t))\} \tilde{\myVec{x}}_{t} + {\rm diag}\{\myVec{c}_{\mathrm{out}}(\sigma(t))\}
\notag\\
&\, \times F_{\myVec{\theta}} \left( c_{\mathrm{in}}(\sigma(t))\tilde{\myVec{x}}_t, c_{\mathrm{noise}}(\sigma(t)), \kappa \right),
\label{eq:masked_edm_preconditioning}
\end{align}
where $c_{\mathrm{in}}(\cdot)$ and $c_{\mathrm{noise}}(\cdot)$ are given in \eqref{eq:standard_edm_coefficients}, while
\begin{align}
\myVec{c}_{\mathrm{skip}}(\sigma) &=
 \frac{\sigma_{\mathrm{data}}^2} {\sigma^2+\sigma_{\mathrm{data}}^2} \bar{\myVec{q}},
\label{eq:masked_cskip}\\
\myVec{c}_{\mathrm{out}}(\sigma)
&=
\frac{\sigma\sigma_{\mathrm{data}}}  {\sqrt{\sigma^2+\sigma_{\mathrm{data}}^2}}\bar{\myVec{q}}  + \sigma_{\mathrm{data}}(\myVec{1}-\bar{\myVec{q}} ).
\label{eq:masked_cout}
\end{align}
For a trusted channel, i.e., $[\bar{\myVec{q}}]_c=1$, these expressions reduce to the standard \ac{edm} coefficients in \eqref{eq:standard_edm_coefficients}. For a masked channel, $[\bar{\myVec{q}}]_c=0$, the skip coefficient vanishes, preventing its pure-noise realization from being directly propagated to the output, while $[\myVec{c}_{\mathrm{out}}]_c=\sigma_{\mathrm{data}}$ provides the appropriate scaling for predicting the missing component through $F_\myVec{\theta}$. The trusted order $\kappa$ is additionally provided explicitly to $F_\myVec{\theta}$, allowing the same network to distinguish between the different order-availability patterns.

Accordingly, the parameters $\myVec{\theta}$ are learned by minimizing the channel-wise preconditioned denoising objective
\begin{align} 
\mathcal{L}(\myVec{\theta}) = \mathbb{E}_{\myVec{x}_0,\myVec{\epsilon},\tau,\kappa} \Bigg[& \frac{1}{C \cdot F \cdot T} \Bigg\|
F_{\myVec{\theta}} \left( c_{\mathrm{in}} (\tau)\tilde{\myVec{x}}_\tau, c_{\mathrm{noise}}(\tau), \kappa \right) \notag 
\\[-0.5ex]
& \quad-
\frac{ \myVec{x}_{0} - {\rm diag}\{\myVec{c}_{\mathrm{skip}}(\tau)\} \tilde{\myVec{x}}_{\tau} }{ \myVec{c}_{\mathrm{out}}(\tau) } \Bigg\|_2^2 \Bigg], 
\label{eq:masked_edm_loss}
\end{align}
where $C=(N_p+1)^2$ is the number of Ambisonics channels, $F$ and $T$ denote the numbers of frequency bins and time frames, respectively,  $\tau$ denotes a sampled diffusion noise level, and vector division is taken element-wise. Through the joint sampling of $\kappa$ and $\tau$, a single array-independent denoiser thus learns both behaviors required at inference: for $\kappa=N_p$ it models the full order-$N_p$ Ambisonics distribution, while for $\kappa<N_p$ it learns to complete the higher-order representation from the trusted lower-order components.

\subsection{Discussion}
\label{ssec:discussion}

The proposed \ac{adeups} construction is tailored to the coupled challenges in Subsection~\ref{sec:ProblemFormulation} by separating the information supplied by the learned Ambisonics prior from that supplied by the physical acquisition model. In particular, rather than relying solely on inversion of the possibly ill-conditioned steering matrix, the posterior trajectory combines a learned prior over ideal Ambisonics with measurement consistency through $\mathcal{A}(\cdot)$. This provides a data-driven regularization mechanism for the noise amplification and non-uniform sensitivity in \ref{item:c1}--\ref{item:c2}, while explicitly accounting for the array-dependent acquisition effects underlying spatial aliasing in \ref{item:c3}. The order-conditioned generation score exploits learned inter-order dependencies to infer components that are insufficiently determined by the measurements, thereby addressing the high-order recovery challenge in~\ref{item:c4}, with the likelihood score maintaining consistency with the available physical observations. Crucially, the diffusion prior is trained only on ideal Ambisonics signals, whereas the steering and encoding matrices enter solely through the test-time forward operator. Consequently, a single learned model can be applied to different microphone geometries and channel counts without retraining,  addressing~\ref{item:c5}.

The inference procedure introduces several parameters governing the relative strength and numerical realization of posterior sampling. The coefficients $\lambda_g$, $\lambda_{r_1}$, and $\lambda_{r_2}$ balance generation and refinement, $\sigma_{\mathrm{th}}$ determines when the sampler transitions to refinement-only operation, and $\eta'$ controls the strength of measurement guidance. The noise schedule, number of sampling steps, and stochastic-Churn parameters further determine the numerical accuracy and computational cost of the reverse trajectory, while the distribution $p_\kappa$ used during training determines the trusted-order configurations to which the denoiser is exposed. These quantities can be selected on held-out validation data and subsequently kept fixed across array configurations. In contrast, the target order $N_p$ is specified by the reconstruction problem rather than being an algorithmic hyperparameter. Its relation to the effective acquisition order $N_{\mathrm{eff}}$ nevertheless affects the fidelity of the forward model: when $N_p<N_{\mathrm{eff}}$, contributions from orders above $N_p$ are omitted from~\eqref{eq:degradation_operator}, resulting in model mismatch due to unmodeled spatial components. The impact of this mismatch is examined empirically in Subsection~\ref{sssec:ablation}.

The formulation also suggests several directions for further development. First, the current trusted order $N_{\mathrm{enc}}$ is determined from the microphone count, whereas in practice observability also depends on frequency, array geometry, and the conditioning of the corresponding steering matrix. Replacing this hard order threshold by frequency- and array-dependent confidence information, potentially through soft order or channel masks, could more accurately distinguish measured from prior-dominated components. Further, the present likelihood assumes that the array steering matrix is accurately known; extending posterior sampling to account for calibration errors or uncertainty in the physical acquisition model would broaden its applicability to practical recording systems. 
These extensions are left for future study.


\section{Empirical Evaluation}
\label{sec:numerical_study}

Here, we numerically evaluate \ac{adeups}\footnote{The source code and hyperparameters used are available at  \url{https://github.com/Amitmils/ADEUPS}}, and compare it with several Ambisonics reconstruction methods. The evaluation is structured as follows: Subsection~\ref{ssec:exp_setup} outlines the experimental setup and the datasets used for training and evaluation. Subsection~\ref{ssec:methodology} details the methodology, including the model configurations, baselines, and the evaluation metrics, while  Subsection~\ref{ssec:numerical_results} reports the numerical results.

\subsection{Experimental Setup}\label{ssec:exp_setup}
\subsubsection{Training Data} 
For training, we use order $N_p = 5$ \acp{arir} sampled at $16\,\text{kHz}$, generated using HARP room simulation code~\cite{saini2025harp}. We randomly generate rooms with dimensions within $[6, 10] \times [6, 10] \times [2, 3]\,\text{m}$ and reverberation times $T_{60} \in [0.1, 0.5]\,\text{s}$. The Ambisonics measurement point is randomly positioned within a $1 \times 1\,\text{m}$ region on the $xy$-plane at a height of $1.2$ to $1.5\,\text{m}$ along the $z$-axis. Point sources are positioned with distances to the Ambisonics measurement point randomly sampled from $0.8$ to $1.5\,\text{m}$, with \acp{doa} also drawn randomly. In total, we generate 2000 rooms with $10$ \acp{arir} per room, yielding $20{,}000$ \acp{arir}. For the speech dataset, we use VCTK~\cite{yamagishi2019cstr}, which consists of approximately $44\,\text{hours}$ of data. We employ an on-the-fly data loader that randomly crops $2.04\,\text{s}$ speech segments per source at runtime and convolves them with the corresponding \acp{arir}. The number of active speakers is randomly chosen to be $1$ or $2$.

\subsubsection{Evaluation Data}\label{ssec:eval_data_framework}
To evaluate the models, we generate $100$ acoustic scenes following the training framework, with two key differences: $(i)$ the Ambisonics order is set to $N_{\text{eff}} = 15$; and $(ii)$ speech signals are taken from the WSJ0 dataset~\cite{garofolo1993csr}. 
We evaluate performance across $13$ test array geometries: the 7-microphone Project Aria glasses~\cite{engel2023project} alongside $4$ irregular realizations for each microphone count $Q \in \{4, 5, 6\}$. For Project Aria, we utilize its measured steering matrix, whereas for the irregular geometries, steering matrices are computed analytically. Following~\cite{heikkinen2025gen}, the microphone positions for the irregular arrays are randomly sampled subject to minimum and maximum pairwise distances of $0.02\,\text{m}$ and $0.18\,\text{m}$, respectively. Microphone signals are generated by projecting the ground-truth Ambisonics onto each array geometry via~\eqref{eq:matrix_signal_acquisition} and adding white Gaussian noise at an \ac{snr} of $50\,\text{dB}$, totaling in $1300$ microphone signals.

\subsection{Methodology}\label{ssec:methodology}
We base our diffusion model  on the NCSN++M~\cite{lemercier2023analysing} backbone. We incorporate a fully connected layer for trusted order conditioning, replace additive conditioning with \acl{adaln}~\cite{peebles2023scalable}, and resize the input and output fully connected layers to $(N_p + 1)^2$ channels, resulting in a model with $31.1\,\text{M}$ parameters.

For the \ac{stft}, an FFT length of 510 samples, a hop size of 128 samples, and a Hann window were employed, resulting in 256 frequency bins and 256 time frames. The compression parameters were set to $\alpha = 0.67$ and $\beta = 3$ to achieve a unit standard deviation ($\sigma_{\text{data}} = 1$). 
During training, the noise schedule is configured with $\sigma_{\max} = 80$, $\sigma_{\min} = 2 \times 10^{-3}$, and $\rho = 10$. The order-masking probability is set to $p(\kappa = N_p) = p(\kappa = 2) = 0.5$. During inference, we adopt the same parameters except for $\sigma_{\max}$, which is reduced to $4$. Sampling is performed with $M = 90$ steps, resulting in $150$ function evaluations, using stochastic churn parameters $S_{\text{churn}} = 5$, $S_{\text{noise}} = 1$, $S_{t\min} = 0$, and $S_{t\max} = 50$. The score weights are set to $\lambda_g = \lambda_{r1} = 0.7$, $\lambda_{r2} = 1$, and $\eta' = 0.5$, with a guidance threshold of $\sigma_{\text{th}} = 0.4$.

\subsubsection{Benchmarks}
We compare our framework against both model-based methods and learning-based methods, with the former being array-agnostic by design. To the best of our knowledge, no existing data-driven  method supports varying array configurations and sizes, and we thus benchmark against both semi-agnostic and array-specific approaches:
\begin{itemize}
\item {\em Linear Encoder}: The  encoder $\myVec{E}_{N_{\text{enc}}}$ defined in \eqref{eq:linear_encoder}.
\item {\em Parametric Method}: Based on \cite{mccormack2022parametric}, we estimate the direct and reverberant components of the sound field and resynthesize the Ambisonics signal up to the target order. To establish an upper bound, we use oracle \acp{doa} to estimate direction of sound sources.
\item {\em UNet-Encoder}: Based on \cite{heikkinen2024neural}, this approach utilizes a U-Net architecture to estimate the encoding matrix $\myVec{E}_{N_{\text{enc}}}$. As it is array-specific, we train it on one of the four-microphone arrays from the test set. The network has a size of 7.9M parameters.
\item {\em Gen-A}: Based on \cite{heikkinen2025gen}, we employ a U-Net \ac{nn} that estimates the encoding matrix conditioned on the microphone positions to encode up to the spatially resolvable order. Because the \ac{nn} is constrained to a fixed channel count, we train it across diverse array geometries with $Q=5$ randomly positioned microphones. This network is of size 10.2M parameters.
\item {\em Diff-Encoder}: We train a diffusion model using the NCSN++M architecture to generate an $N_p$-order Ambisonics signal conditioned on the captured microphone signals $\myVec{p}(k)$. Conditioning is performed by concatenating $\myVec{p}(k)$ with the corrupted state $\myVec{x}_t$ along the channel dimension. This approach is array-specific, and is  trained on one of the four-microphone arrays from the test set. The network size in this case is 30.8M parameters.
\item {\em Diff-Upscaler}: Inspired by \cite{milstein2026generativelearningambisonicupscaling}, we utilize the NCSN++M architecture to upscale from \ac{foa} to order $N_p$. We train the model on \ac{foa} signals linearly encoded from randomly positioned arrays with $Q \in \{4,6\}$ microphones. This network is of size 30.8M parameters.
\end{itemize}
A summary of the models used in the evaluation is presented in Table \ref{tab:models_summary}.
\begin{table*}[t]
\centering
\caption{Summary of Models and their Specifications.}
\label{tab:models_summary}
\small
\begin{tabular}{lcccc}
\toprule
\textbf{Method} & \textbf{Parameters} & \textbf{Input Signal} & \textbf{Output Order} & \textbf{Supported Arrays} \\
\midrule
Linear Encoder       & 0     & Microphones & $N_{enc}$     & Arbitrary \\
Parametric    & 0     & Microphones & $N_p$ & Arbitrary \\
UNet-Encoder         & 7.9M  & Microphones & $N_{enc}$     & Fixed geometry ($Q=4$) \\
Gen-A                & 10.2M & Microphones & $N_{enc}$     & Arbitrary geometry ($Q=5$) \\
Diff-Encoder         & 30.8M & Microphones & $N_p$ & Fixed geometry ($Q=4$) \\
Diff-Upscaler        & 30.8M & \ac{foa}    & $N_p$ & Arbitrary \\
\ac{adeups} (ours) & 31.1M & Microphones & $N_p$ & Arbitrary  \\
\bottomrule
\end{tabular}
\vspace{0.5cm}
\end{table*}

\subsubsection{Evaluation Metrics}
Let $\myVec{z}$ and $\hat{\myVec{z}}$ denote the reference and reconstructed time-domain Ambisonic signals, respectively. Furthermore, let $Z_c(t, f)$ and $\hat{Z}_c(t, f)$ represent their uncompressed \ac{stft} coefficients for channel $c \in \{1, \dots, C\}$, with $t$ and $f$ denoting the time-frame and frequency-bin indices, respectively. We evaluate performance using the following metrics:
\\
\noindent
    $(i)$ \textbf{Log-Magnitude Spectral Error:} Measures the spectral magnitude fidelity across all channels $C$, time frames $T$, and frequency bins $F$ as
    \begin{equation*}
        {\rm LogSE} \!=\! \frac{1}{CTF} \sum_{c=1}^{C} \sum_{t=1}^{T} \sum_{f=1}^{F} \left| 20\cdot \log_{10} \left(\frac{|Z_c(t, f)|}{|\hat{Z}_c(t, f)|}\right)\!   \right|,
    \end{equation*}\\
\noindent
    $(ii)$   \textbf{Mean Spectral Coherence:} Evaluates the average normalized cross-spectral coherence between the reference and estimated signals:
    \begin{equation*}
       {\rm MSC} = \frac{1}{CF} \sum_{c=1}^{C} \sum_{f=1}^{F} \frac{\left| \sum_{t=1}^{T} Z_c^*(t, f) \hat{Z}_c(t, f) \right|}{\sqrt{\sum_{t=1}^{T} |Z_c(t, f)|^2} \sqrt{\sum_{t=1}^{T} |\hat{Z}_c(t, f)|^2}},
    \end{equation*}
\noindent
    $(iii)$  \textbf{\ac{sisdr}:} Evaluates overall time-domain fidelity while remaining invariant to global gain mismatches via a scalar projection $\alpha \in \mathbb{R}$
\begin{equation}
        \text{\acs{sisdr}} = 10 \log_{10} \left( \frac{\|\alpha \myVec{z}\|^2}{\|\myVec{\hat{z}} - \alpha \myVec{z}\|^2} \right), \quad \text{with } \alpha = \frac{|\langle\myVec{\hat{z}}, \myVec{z}\rangle|}{\|\myVec{z}\|^2},
    \end{equation}
 
\subsection{Results} 
We present the results in three parts. First, we compare the proposed method to existing array-agnostic baselines, followed by a comparison against array-specific methods. Finally, we conduct an ablation study to analyze our key design choices.
\label{ssec:numerical_results}
\subsubsection{Full- and Semi-Agnostic Methods Evaluation}
Table~\ref{tab:hoa_order_comparison} presents performance of the array-agnostic methods across different orders and array configurations. 
For first-order encoding, we compare both model-based approaches with \ac{adeups} and Gen-A (which is restricted to five-microphone arrays). Our method outperforms all baselines in SI-SDR by a substantial margin and achieves superior spectral error across nearly all configurations, while remaining second-best in coherence.

For orders 2-5, the linear encoder and Gen-A are inapplicable; thus, we also evaluate Diff-Upscaler, which operates on linearly encoded \ac{foa}. Because its input lies in the Ambisonics domain rather than the microphone domain, it is array-agnostic in the sense that varying microphone counts do not necessitate retraining. We trained Diff-Upscaler exclusively on four- and six-microphone arrays to evaluate its behavior on two out-of-distribution scenarios. On \ac{foa} derived from a five-microphone array, the upscaler performs comparably to the four- and six-microphone cases; this demonstrates that although it was not trained on these specific spatial aliasing artifacts, they closely resemble those of the training arrays, which share an open-sphere topology. Conversely, on the Aria glasses array, the upscaler fails entirely. This highlights a fundamental pitfall of supervised training on specific array configurations: generalization becomes unreliable for geometries that deviate significantly from the training set. 

These results highlight the strength of \ac{adeups}, which achieves superior \acs{sisdr} across all array configurations without having observed any of them during training. This performance gap between diffusion models of identical capacity stems from the underlying distribution being learned: \ac{adeups} trains on a clean $N_p$-order prior and incorporates array-specific forward operators solely during inference, thereby simplifying optimization. In contrast, the diffusion upscaler must learn a joint distribution entangled with diverse encoding artifacts, substantially increasing training complexity and hindering generalizability.
\begin{table*}[htbp]
\centering
\caption{Performance Comparison Across Orders and Arrays (bold indicates best performance).}
\label{tab:hoa_order_comparison}
\resizebox{\textwidth}{!}{%
\begin{tabular}{cc|cccc|cccc|cccc}
\toprule
\multirow{2}{*}{\textbf{Order}} & \multirow{2}{*}{\textbf{Method}} & \multicolumn{4}{c|}{\textbf{\acs{sisdr} [dB] $\uparrow$}} & \multicolumn{4}{c|}{\textbf{Mean Spec. Error [dB] $\downarrow$}} & \multicolumn{4}{c}{\textbf{Coherence $\uparrow$}} \\
\cmidrule(lr){3-6} \cmidrule(lr){7-10} \cmidrule(lr){11-14}
 & & \textbf{4 Mics} & \textbf{5 Mics} & \textbf{6 Mics} & \textbf{Aria} & \textbf{4 Mics} & \textbf{5 Mics} & \textbf{6 Mics} & \textbf{Aria} & \textbf{4 Mics} & \textbf{5 Mics} & \textbf{6 Mics} & \textbf{Aria} \\
\midrule
\multirow{4}{*}{1} 
 & Linear          & 6.54           & 7.63           & 8.86           & 3.40          & 9.50          & 9.28          & 8.33          & 10.9          & \textbf{0.82} & \textbf{0.82} & \textbf{0.83} & \textbf{0.76} \\
 & Param. (oracle) & 3.73           & 3.67           & 3.77           & 2.28          & 5.83          & 5.87          & 5.85          & \textbf{6.22} & 0.71          & 0.70          & 0.71          & 0.67          \\
 & Gen-A           & -              & 3.75           & -              & -             & -             & 7.37          & -             & -             & -             & 0.78          & -             & -             \\
 & ADEUPS          & \textbf{10.52} & \textbf{11.27} & \textbf{13.13} & \textbf{6.44} & \textbf{5.61} & \textbf{5.65} & \textbf{5.15} & 7.42          & 0.79          & 0.79          & \textbf{0.83} & 0.66          \\
\midrule
\multirow{3}{*}{2} 
 & Param. (oracle) & 2.41          & 2.55          & 2.60           & 1.50          & \textbf{6.27} & \textbf{6.25} & 6.23          & \textbf{6.53} & 0.69          & 0.68          & 0.68          & 0.64          \\
 & Diff-Upscaler   & 5.71          & 6.19          & 6.74           & -3.36         & 9.20          & 9.03          & 8.40          & 20.18         & 0.74          & 0.74          & 0.76          & 0.63          \\
 & ADEUPS          & \textbf{8.34} & \textbf{9.49} & \textbf{11.04} & \textbf{5.87} & 6.33          & 6.28          & \textbf{5.84} & 7.86          & \textbf{0.76} & \textbf{0.77} & \textbf{0.81} & \textbf{0.65} \\
\midrule
\multirow{3}{*}{3} 
 & Param. (oracle) & 1.80          & 2.00          & 2.00          & 1.00          & \textbf{6.47} & \textbf{6.44} & 6.41          & \textbf{6.69} & 0.67          & 0.66          & 0.67          & \textbf{0.63} \\
 & Diff-Upscaler   & 5.30          & 5.70          & 6.13          & -6.60         & 9.18          & 9.02          & 8.50          & 20.95         & 0.70          & 0.70          & 0.72          & \textbf{0.63} \\
 & ADEUPS          & \textbf{7.58} & \textbf{8.59} & \textbf{9.69} & \textbf{5.29} & 6.67          & 6.62          & \textbf{6.23} & 8.09          & \textbf{0.74} & \textbf{0.75} & \textbf{0.79} & 0.60          \\
\midrule
\multirow{3}{*}{4} 
 & Param. (oracle) & 1.42          & 1.62          & 1.64          & 0.70          & \textbf{6.65} & \textbf{6.61} & \textbf{6.58} & \textbf{6.82} & 0.66          & 0.65          & 0.66          & \textbf{0.62} \\
 & Diff-Upscaler   & 4.79          & 5.26          & 5.64          & -9.13         & 9.21          & 9.05          & 8.60          & 20.26         & 0.68          & 0.68          & 0.69          & 0.54          \\
 & ADEUPS          & \textbf{6.99} & \textbf{7.94} & \textbf{8.82} & \textbf{4.70} & 6.99          & 6.95          & \textbf{6.58} & 8.33          & \textbf{0.73} & \textbf{0.74} & \textbf{0.77} & \textbf{0.62} \\
\midrule
\multirow{3}{*}{5} 
 & Param. (oracle) & 1.78          & 1.39          & 1.40          & 0.50          & \textbf{6.77} & \textbf{6.71} & \textbf{6.67} & \textbf{6.91} & 0.65          & 0.64          & 0.65          & \textbf{0.61} \\
 & Diff-Upscaler   & 4.22          & 4.82          & 5.18          & -11.23        & 9.26          & 9.09          & 8.67          & 18.82         & 0.65          & 0.65          & 0.67          & 0.52          \\
 & ADEUPS          & \textbf{6.34} & \textbf{7.23} & \textbf{8.00} & \textbf{4.02} & 7.24          & 7.20          & 7.42          & 8.52          & \textbf{0.71} & \textbf{0.72} & \textbf{0.76} & \textbf{0.61} \\
\bottomrule
\end{tabular}%
}
\end{table*}

\subsubsection{Array Specific Methods Evaluation}
We next evaluate \ac{adeups} against array-specific baselines: Diff-Encoder, which directly maps microphone signals to $N_p=5$ order Ambisonics signals, and  UNet-Encoder, which estimates an encoding matrix to project the microphone signals to $N_{\text{enc}}=1$ order Ambisonics. The evaluation results on a four-microphone\footnote{Microphone positions in meters: $(-0.08, 0.01, -0.05)$, $(0.01, 0.00, 0.05)$, $(0.02, -0.01, 0.05)$, and $(0.03, -0.00, -0.05)$.\label{fn:4_mic_array}} open array are summarized in Table~\ref{tab:array_specific_hoa_order_comparison}.

\begin{table}[htbp]
\centering
\caption{Performance Comparison for a Specific 4 Mic Array\textsuperscript{\ref{fn:4_mic_array}}.}
\label{tab:array_specific_hoa_order_comparison}
\resizebox{\columnwidth}{!}{%
\begin{tabular}{c l ccc}
\toprule
\textbf{Order} & \textbf{Method} & \textbf{\acs{sisdr} [dB]} $\uparrow$ & \textbf{MSE [dB]} $\downarrow$ & \textbf{Coherence} $\uparrow$ \\
\midrule
\multirow{3}{*}{1} 
 & UNet-Encoder  & 6.65           & \textbf{5.73} & \textbf{0.85} \\
 & Diff-Encoder  & 9.60           & 5.98          & 0.78 \\
 & ADEUPS        & \textbf{10.83} & 5.81          & 0.77 \\
\midrule
\multirow{2}{*}{2} 
 & Diff-Encoder  & 7.60           & \textbf{6.34} & 0.74 \\
 & ADEUPS        & \textbf{8.42}  & 6.45          & \textbf{0.76} \\
\midrule
\multirow{2}{*}{3} 
 & Diff-Encoder  & 7.03           & \textbf{6.57} & 0.72 \\
 & ADEUPS        & \textbf{7.72}  & 6.75          & \textbf{0.74} \\
\midrule
\multirow{2}{*}{4} 
 & Diff-Encoder  & 6.52           & \textbf{6.78} & 0.70 \\
 & ADEUPS        & \textbf{7.17}  & 7.05          & \textbf{0.72} \\
\midrule
\multirow{2}{*}{5} 
 & Diff-Encoder  & 6.01           & \textbf{7.00} & 0.68 \\
 & ADEUPS        & \textbf{6.52}  & 7.27          & \textbf{0.71} \\
\bottomrule
\end{tabular}%
}
\end{table}

For first-order Ambisonics, our method outperforms both baselines in \acs{sisdr}, whereas the UNet-Encoder achieves the best performance in terms of coherence and spectral error. For orders 2 through 5, we compare directly with the Diff-Encoder; our method achieves higher \acs{sisdr} and comparable coherence, while underperforming the baseline by only a small margin in spectral error. 
Overall, these results demonstrate the robustness and flexibility of our approach compared to methods trained on a specific geometry. Crucially, when evaluated on an unseen 4-microphone configuration, the baselines fail entirely, whereas our method maintains consistent performance.

\subsubsection{Ablation Study}
\label{sssec:ablation}
To quantify the individual contributions of our proposed design choices, we conduct an ablation study on the evaluation dataset, reporting the mean metrics across all microphone arrays for each Ambisonics order. Table~\ref{tab:ablation_added_design} summarizes the ablation sequence where the sampling steps where chosen such that each method had 150 model evaluations.

\begin{table*}[htbp]
\centering
\caption{\small Ablation study evaluating the incremental contributions of order-masking and score-blending. The baseline configuration corresponds to \acs{adeps}~\cite{milstein2026array} (without masking or conditioning, using a first-order stochastic Euler sampler). All configurations are evaluated under an identical budget of 150 model evaluations. Best results are shown in \textbf{bold}.}
\label{tab:ablation_added_design}
\resizebox{\textwidth}{!}{%
\begin{tabular}{c|ccccc|ccccc|ccccc}
\toprule
\multirow{2}{*}{\raisebox{0.4ex}{\textbf{Method}}\textbf{/}\raisebox{-0.4ex}{\textbf{Order}}}& \multicolumn{5}{c|}{\textbf{\acs{sisdr} [dB] $\uparrow$}} & \multicolumn{5}{c|}{\textbf{Mean Spec. Error [dB] $\downarrow$}} & \multicolumn{5}{c}{\textbf{Coherence $\uparrow$}} \\
\cmidrule(lr){2-6} \cmidrule(lr){7-11} \cmidrule(lr){12-16}
 & \textbf{1} & \textbf{2} & \textbf{3} & \textbf{4} & \textbf{5} & \textbf{1} & \textbf{2} & \textbf{3} & \textbf{4} & \textbf{5} & \textbf{1} & \textbf{2} & \textbf{3} & \textbf{4} & \textbf{5} \\
\midrule
ADEPS & 10.53 & 5.79 & 3.87 & 2.52 & 1.30 & 5.48 & 6.57 & 7.47 & 8.31 & 9.16 & 0.76 & 0.72 & 0.69 & 0.66 & 0.64  \\
ADEPS + Masking & 10.38 & 7.47 & 5.91 & 4.64 & 3.34 & 5.52 & 6.24 & 6.77 & 7.30 & 7.93 & 0.77 & 0.74 & 0.71 & 0.69 & 0.67 \\
ADEPS + Masking + Ref Cond. & 10.58 & 7.29 & 5.65 & 4.39 & 3.20 & \textbf{5.42} & \textbf{6.19} & 6.79 & 7.36 & 8.00 & \textbf{0.79} & 0.75 & 0.73 & 0.70 & 0.68 \\
ADEPS + Masking + Gen Cond. & 0.40 & 2.70 & 3.88 & 4.40 & 4.47 & 12.89 & 10.09 & 9.12 & 8.90 & 8.83 & 0.59 & 0.68 & 0.70 & 0.71 & 0.70 \\
ADEPS + Masking + Score Blending & 10.04 & 4.40 & 3.04 & 2.62 & 1.70 & 6.68 & 10.24 & 11.60 & 12.58 & 13.05 & 0.75 & 0.73 & 0.71 & 0.70 & 0.68 \\
ADEPS + Masking + Score Blending + Scheduler (ADEUPS) & \textbf{11.24} & \textbf{9.33} & \textbf{8.37} & \textbf{7.67} & \textbf{6.95} & 5.61 & 6.28 & \textbf{6.62} & \textbf{6.95} & \textbf{7.20} & \textbf{0.79} & \textbf{0.77} & \textbf{0.75} & \textbf{0.74} & \textbf{0.72}  \\
\bottomrule
\end{tabular}%
}
\end{table*}

We define the baseline as our previous framework, \acs{adeps}~\cite{milstein2026array}, which performs Ambisonics encoding via \ac{dps} using an EDM-based learned prior with a first-order stochastic Euler sampler, omitting both order-masking during training and score-blending during inference. Incorporating \ac{hoa} masking into training---even without conditioning---substantially improves the \acs{sisdr} for higher-order components while maintaining comparable performance on lower-order ones. 

Next, we evaluate models trained with both masking and conditioning, examining the individual effects of conditioning on either the \textit{refinement} score or the \textit{generation} score. Conditioning solely on the refinement score yields results similar to the masking-only framework. In contrast, using only the generation score severely degrades performance at lower orders due to the absence of a skip connection. Notably, unlike the other configurations, the generation-only model exhibits a reversed performance trend, where performance improves with higher Ambisonics orders.

\begin{figure}
    \centering
    \includegraphics[width=1\linewidth]{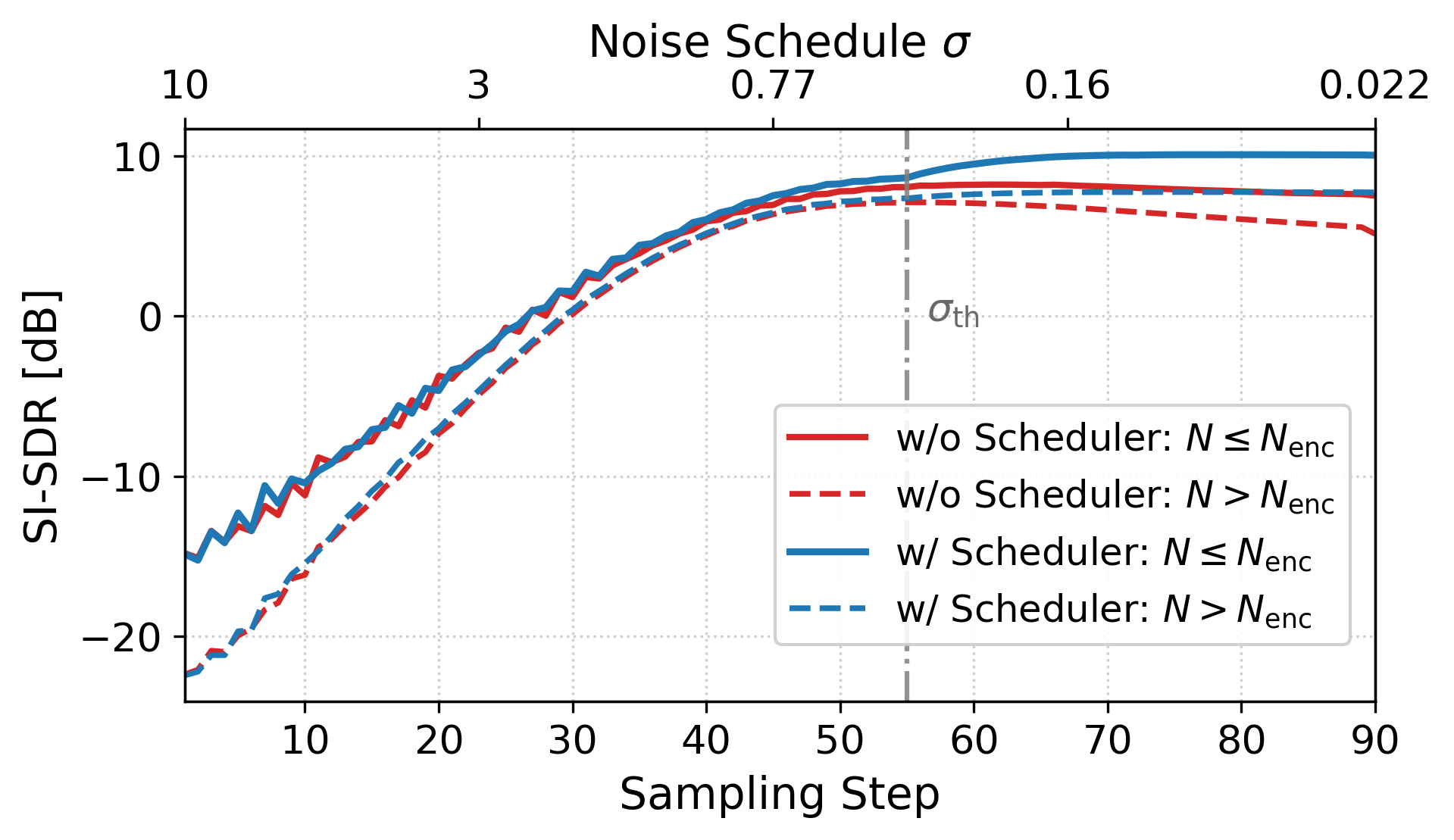}
    \caption{\acs{sisdr} over time for \ac{adeups} with and without scheduler.}
    \label{fig:iter_plot}
\end{figure}

This observation motivates a hybrid approach that leverages the strong low-order performance of the refinement score while utilizing the low-order components to guide higher-order generation. However, naive score blending initially yields suboptimal results because the generation score degrades quality at lower noise levels. At these noise levels, high-order components are already established; attempting to generate them primarily from low-order components becomes ineffective and even detrimental. To address this, we empirically define a threshold $\sigma_{\text{th}}$, below which the sampler transitions from score blending exclusively to the refinement score. Figure~\ref{fig:iter_plot} illustrates the iteration-wise dynamics with and without the scheduler---evaluated on the average SI-SDR for both low-order and high-order components---demonstrating its effectiveness in stabilizing convergence during sampling.

\begin{figure}
    \centering
    \includegraphics[width=1\linewidth]{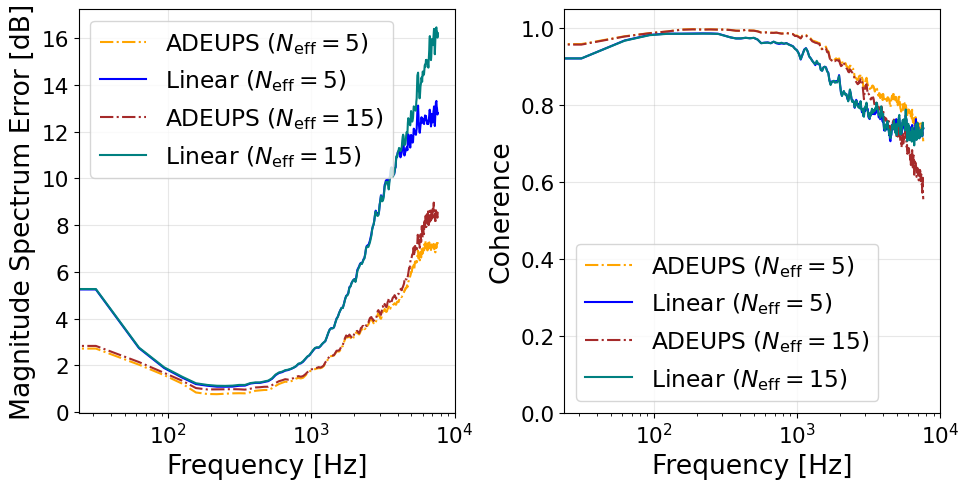}
    \caption{Spectral error and coherence for truncation orders $N_{\text{eff}} = 5$ and $N_{\text{eff}} = 15$.}
    \label{fig:spectral_analysis}
\end{figure}

We next compare performance across different values of $N_{\text{eff}}$, evaluating our test set at $N_{\text{eff}} \in \{5, 10\}$. Table~\ref{tab:perf_per_effective_order} reports the results per Ambisonics order, averaged across all microphone arrays.
At $N_{\text{eff}} = 5$, we observe peak performance because $N_p = N_{\text{eff}}$, rendering our forward measurement model~\eqref{eq:degradation_operator} exact and thus establishing the upper bound on performance. Beyond this, we observe a noticeable performance drop from $N_{\text{eff}} = 5$ to $N_{\text{eff}} = 10$, but only a minimal change between $N_{\text{eff}} = 10$ and $N_{\text{eff}} = 15$. These findings confirm two key insights: $(i)$ The initial performance degradation stems from the truncated prior order ($N_p = 5$). Because the prior cannot represent components beyond $N_p$, it fails to account for energy folding caused by spatial aliasing from higher orders. This behavior is further corroborated by the spectral analysis in Fig.~\ref{fig:spectral_analysis}, which reveals that the performance discrepancy between $N_{\text{eff}} = 5$ and $N_{\text{eff}} = 15$ is concentrated at high frequencies where spatial aliasing is most pronounced; improving performance under such conditions would thus require a higher-order prior. $(ii)$ The marginal difference between $N_{\text{eff}} = 10$ and $N_{\text{eff}} = 15$ demonstrates that energy decays rapidly as the order increases. This indicates that $N_{\text{eff}} = 15$ serves as a valid practical approximation of the infinite sum for speech data, supporting the core assumption of our framework.

\begin{table*}[htbp]
\centering
\caption{Effect of $N_{\text{eff}}$ on Performance per Order.}
\label{tab:perf_per_effective_order}
\resizebox{\textwidth}{!}{%
\begin{tabular}{c|ccccc|ccccc|ccccc}
\toprule
\multirow{2}{*}{\textbf{$N_{\text{eff}}$}} & \multicolumn{5}{c|}{\textbf{\acs{sisdr} [dB] $\uparrow$}} & \multicolumn{5}{c|}{\textbf{Mean Spec. Error [dB] $\downarrow$}} & \multicolumn{5}{c}{\textbf{Coherence $\uparrow$}} \\
\cmidrule(lr){2-6} \cmidrule(lr){7-11} \cmidrule(lr){12-16}
 & \textbf{1} & \textbf{2} & \textbf{3} & \textbf{4} & \textbf{5} & \textbf{1} & \textbf{2} & \textbf{3} & \textbf{4} & \textbf{5} & \textbf{1} & \textbf{2} & \textbf{3} & \textbf{4} & \textbf{5} \\
\midrule
5 & \textbf{14.68} & \textbf{11.27} & \textbf{9.92} & \textbf{9.12} & \textbf{8.37} & \textbf{5.05} & \textbf{5.77} & \textbf{6.14} & \textbf{6.47} & \textbf{6.73} & \textbf{0.85} & \textbf{0.83} & \textbf{0.81} & \textbf{0.80} & \textbf{0.78} \\
10 & 11.29 & 9.38 & 8.41 & 7.73 & 7.01 & 5.60 & 6.27 & 6.62 & 6.95 & 7.20 & 0.80 & 0.78 & 0.76 & 0.74 & 0.72 \\
15 & 11.24 & 9.33 & 8.37 & 7.67 & 6.95 & 5.61 & 6.28 & 6.62 & 6.95 & 7.20 & 0.79 & 0.77 & 0.75 & 0.74 & 0.72\\
\bottomrule
\end{tabular}%
}
\end{table*}

\section{Listening Experiment}\label{sec:listening_exp}
While Section~\ref{sec:numerical_study} provides objective evaluations, subjective listening experiments remain essential to capture perceptual attributes beyond the scope of numerical metrics. The experimental setup is detailed in Section~\ref{ssec:listening_setup}, followed by the evaluation methodology in Section~\ref{ssec:listening_methodology}. Finally, the subjective results are discussed in Section~\ref{ssec:listening_results}.

\subsection{Setup}\label{ssec:listening_setup}
We follow the framework outlined in Section~\ref{ssec:eval_data_framework} to generate the evaluation data, setting the dry signal duration to 5~seconds. The listening test evaluates three screen configurations across distinct acoustic and array setups:
\begin{enumerate}[label=$S_\arabic*$, leftmargin=*, series=challenges]
    \item \label{item:s1} \textbf{Single-speaker \ac{foa}}: Room dimensions $6.8 \times 8.2 \times 3.0\,\mathrm{m}$ ($\mathrm{RT}_{60} = 380\,\mathrm{ms}$, $\mathrm{DRR} = -0.4\,\mathrm{dB}$), using a 4-microphone array centered at $(2.9, 6.2, 1.7)\,\mathrm{m}$. A male speaker is positioned at a distance of $1.3\,\mathrm{m}$ with $(\theta, \phi) = (90^\circ, 22^\circ)$.
    \item \label{item:s2} \textbf{Dual-speaker \ac{foa}}: Same acoustic environment as \ref{item:s1} ($\mathrm{RT}_{60} = 380\,\mathrm{ms}$, $\mathrm{DRR} = -0.4\,\mathrm{dB}$), using a 5-microphone array centered at $(2.9, 6.2, 1.7)\,\mathrm{m}$. Two sources are positioned in the room: a male speaker located at $1.3\,\mathrm{m}$ with $(81^\circ, 45^\circ)$ and a female speaker at $1.2\,\mathrm{m}$ with $(110^\circ, -45^\circ)$.
    \item \label{item:s3} \textbf{Single-speaker \ac{hoa}4}: Room dimensions $8.8 \times 8.2 \times 3.0\,\mathrm{m}$ ($\mathrm{RT}_{60} = 450\,\mathrm{ms}$, $\mathrm{DRR} = 3.2\,\mathrm{dB}$), using a 6-microphone array centered at $(3.8, 6.2, 1.7)\,\mathrm{m}$. A female speaker is positioned at a distance of $1.3\,\mathrm{m}$ with $(96^\circ, -32^\circ)$.
\end{enumerate}
For each scenario, four stimuli are generated: (i)~the ground-truth Ambisonic signal at the target order (\ac{foa} for \ref{item:s1}--\ref{item:s2}, \ac{hoa} of order 4 for \ref{item:s3}); (ii)~a linearly encoded \ac{foa} signal; (iii)~the target-order Ambisonics signal produced by \ac{adeups}; and (iv)~an Ambisonics signal generated by a data-driven baseline. Because existing deep learning baselines cannot simultaneously accommodate arbitrary array geometries and higher target orders, different models are used across scenarios: UNet-Encoder and Gen-A (which take raw microphone signals as input) for \ref{item:s1} and \ref{item:s2}, and Diff-Upscaler (which operates on the linearly encoded \ac{foa}) for \ref{item:s3}.

\subsection{Methodology}\label{ssec:listening_methodology}
The subjective evaluation was conducted following the \ac{mushra} methodology~\cite{series2014method}. Participants were instructed to rate the \textit{overall quality} of the stimuli relative to an explicit known reference on a continuous scale from 0 to 100, where 100 denotes no perceptible difference from the reference and lower scores indicate increasing degradation. In each scenario, the ground-truth Ambisonic signal served as the reference (and hidden reference), while the linearly encoded \ac{foa} signal served as a low-anchor.

A total of 11 listeners participated in the experiment. Prior to testing, each subject completed a brief training session to familiarize themselves with the range of sound quality and the test interface. The presentation order of the test screens and stimuli within each screen was randomized to mitigate potential bias.

\subsection{Results}\label{ssec:listening_results}

\begin{figure*}[t]
    \centering
    \begin{subfigure}{0.32\linewidth}
        \centering
        \includegraphics[width=\linewidth]{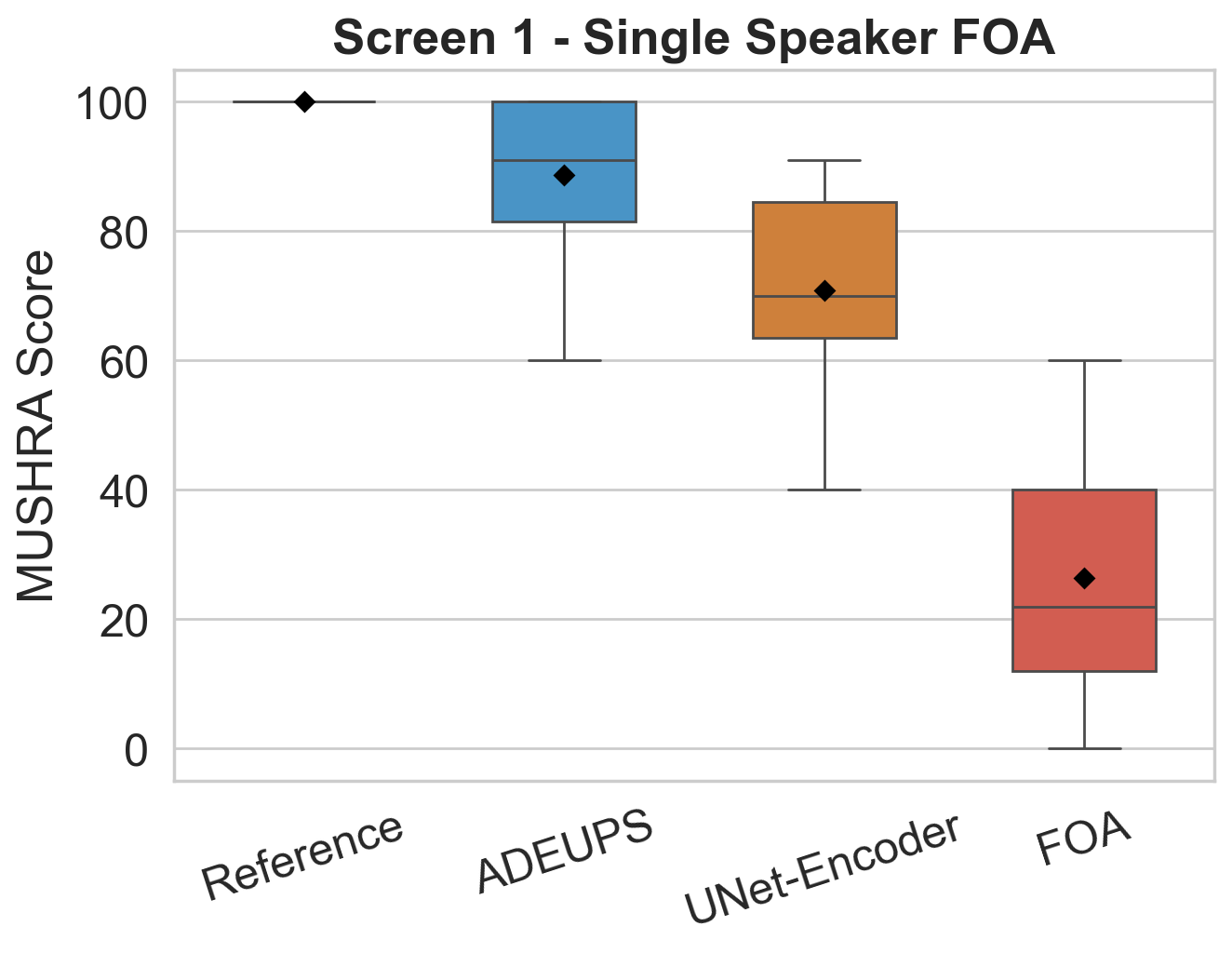}
        \label{fig:mushra_plot_1}
    \end{subfigure}
    \hfill 
    \begin{subfigure}{0.32\linewidth}
        \centering
        \includegraphics[width=\linewidth]{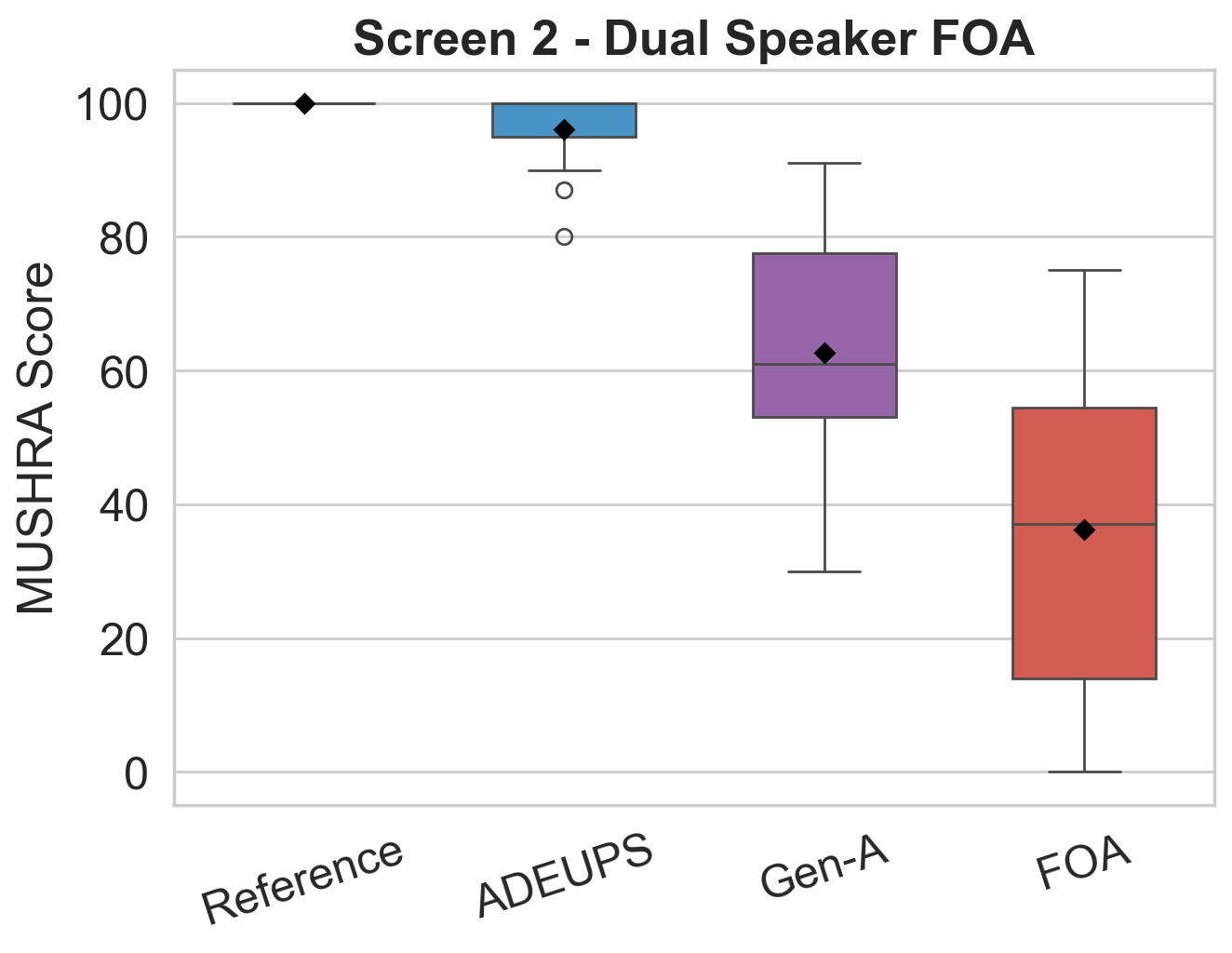}
        \label{fig:mushra_plot_2}
    \end{subfigure}
    \hfill
    \begin{subfigure}{0.32\linewidth}
        \centering
        \includegraphics[width=\linewidth]{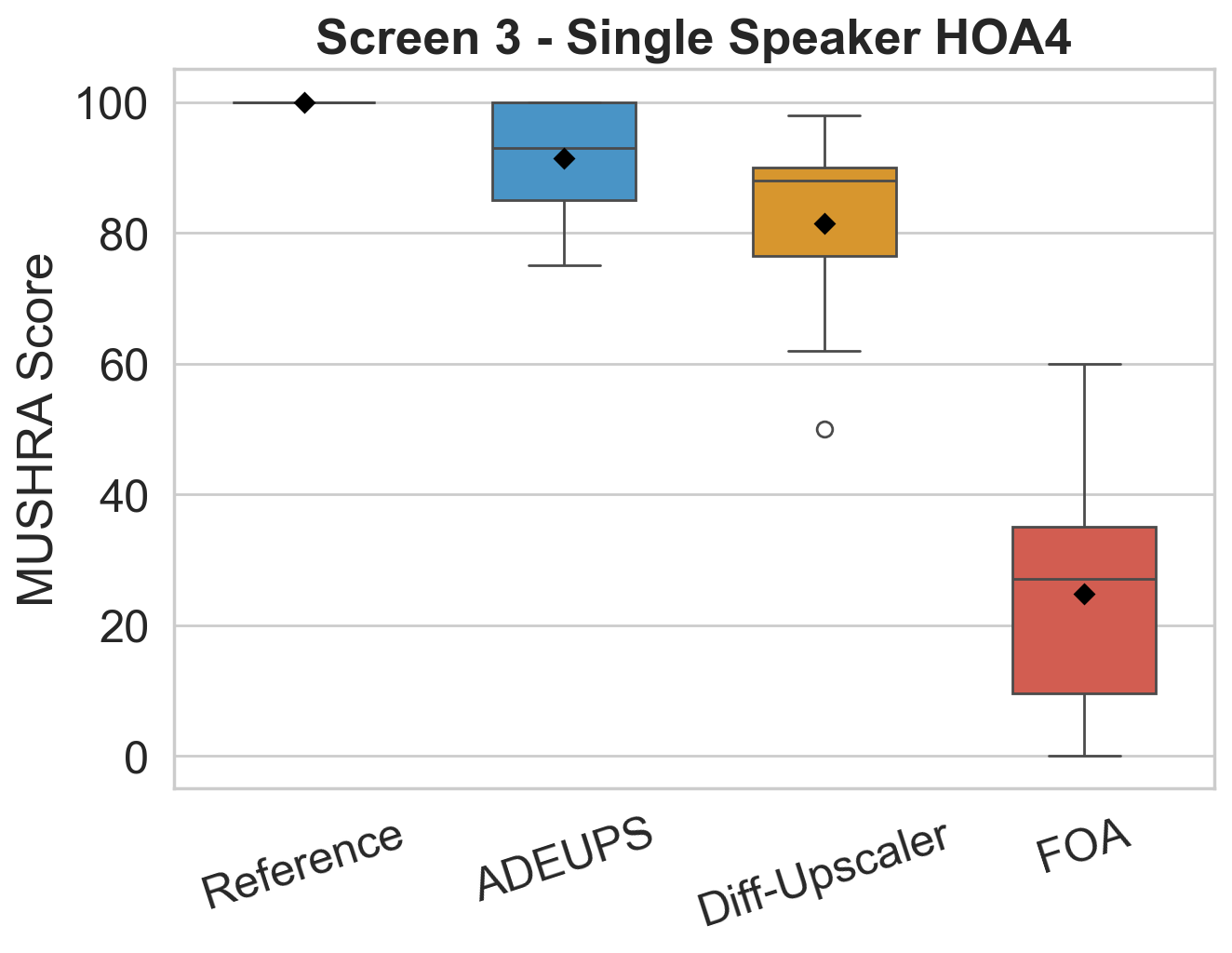}
        \label{fig:mushra_plot_3}
    \end{subfigure}
    \vspace{6pt} 
    \caption{MUSHRA results across test screens: Screen 1 (4-microphone array), Screen 2 (5-microphone array), and Screen 3 (6-microphone array). Horizontal lines denote medians, and diamonds denote means.}
    \label{fig:combined_mushra_plots} 
\end{figure*}

The \ac{mushra} test results are illustrated in Figure~\ref{fig:combined_mushra_plots}. For each test screen, a two-way repeated-measures \ac{anova} was conducted to evaluate the impact of the processing method on subjective quality ratings.

\paragraph*{Screen 1 (\ref{item:s1})} The analysis revealed a statistically significant main effect for the processing method ($F(3, 30) = 22.31, p < .001, \eta_p^2 = 0.64$), indicating that method choice substantially influenced perceived quality. Post-hoc pairwise comparisons with Bonferroni corrections were performed across the \ac{adeups}, ground-truth \ac{foa}, and UNet-Encoder conditions, which achieved mean scores of 100, 88.72, and 70.90, respectively. No statistically significant difference was observed between \ac{adeups} and either the ground-truth signal ($p = 0.10$, $g = 1.16$) or the UNet-Encoder ($p = 0.08$, $g = 1.20$). However, ratings for the ground-truth signal were significantly higher than those for the UNet-Encoder ($p < 0.05$, $g = 2.60$).

\paragraph*{Screen 2 (\ref{item:s2})} Mauchly’s test indicated a violation of sphericity ($\chi^2(5) = 15.35, p = .009$), requiring Greenhouse-Geisser corrections to the degrees of freedom. The main effect of the processing method remained statistically significant ($F(1.88, 18.78) = 40.26, p < .001, \eta_p^2 = 0.72$). Mean scores for \ac{adeups}, ground-truth \ac{foa}, and the UNet-Encoder were 100, 96.09, and 62.68, respectively. Post-hoc Bonferroni-corrected comparisons revealed no significant difference between \ac{adeups} and the ground-truth signal ($p = 0.58$, $g = 0.75$), whereas Gen-A performed significantly worse than both \ac{adeups} ($p < 0.05$, $g = -2.17$) and the ground truth ($p < 0.05$, $g = -2.58$).

\paragraph*{Screen 3 (\ref{item:s3})} Similarly, a highly significant main effect of the processing method was observed ($F(3, 30) = 84.94, p < .001, \eta_p^2 = 0.84$). The mean scores for \ac{adeups}, ground-truth \ac{foa}, and the Diff-Upscaler were 100, 91.45, and 81.45, respectively. Post-hoc comparisons with Bonferroni correction showed no statistically significant difference between \ac{adeups} and either the ground-truth signal or the Diff-Upscaler; however, the ground-truth signal rated significantly higher than the Diff-Upscaler.

\section{Conclusions}
\label{sec: Conclusions}

In this work, we introduced \ac{adeups}, a novel, array-agnostic framework for Ambisonic encoding and spatial upscaling. By formulating the reconstruction task as a diffusion-based inverse problem solved via \ac{dps}, our model learns an acquisition-independent sound-field prior of ideal \ac{hoa} signals. At inference, array-specific configurations are incorporated into the forward operator to invert device-induced physical artifacts. To further enhance reconstruction quality, we introduced two key innovations: order-masking during training to enforce interchannel dependencies, and score-blending during inference to guide sampling trajectory. Objective and subjective evaluations demonstrate that \ac{adeups} achieves performance comparable to—and in many cases exceeding—both array-agnostic model-based baselines and data-driven methods trained on specific array topologies. Crucially, \ac{adeups} operates in a zero-shot manner, requiring no exposure to specific array geometries during training. Future work will explore real-time algorithmic optimizations and extend the framework to dynamic acoustic scenes.


 

\begingroup
{\footnotesize 
\bibliographystyle{IEEEtran}
\bibliography{mybib}}
\endgroup

\end{document}